\documentclass{vgtc}                          % final (conference style)
\graphicspath{{figures/}{pictures/}{images/}{./}} % where to search for the images

\usepackage{times}                     % we use Times as the main font
\usepackage{tabu}                      % only used for the table example
\usepackage{booktabs}                  % only used for the table example
\usepackage{lipsum}                    % used to generate placeholder text
\usepackage{mwe}                       % used to generate placeholder figures

\usepackage{mathcomp}
\usepackage{here}
\usepackage{enumitem}
\usepackage{comment}
\usepackage{color}

\usepackage{mathptmx}                  % use matching math font

\usepackage{siunitx}
\usepackage{ifthen}
\definecolor{pink}{rgb}{0.8, 0.0, 0.6}
\definecolor{blue}{rgb}{0.1, 0.2, 0.8}
\definecolor{red}{rgb}{0.8, 0.1, 0.1}
\definecolor{yellow}{rgb}{0.8, 0.6, 0.0}

\newcommand{\etal}{\textit{et al.}}
\newcommand{\eg}{\textit{e.g.}}

\newboolean{highlight}

\ifthenelse{\boolean{highlight}}{%
    \newcommand{\add}[1]{\textcolor{blue}{#1}}
    \newcommand{\del}[1]{\textcolor{red}{[delete: #1]}}
    \newcommand{\com}[1]{\textcolor{yellow}{[comment: #1]}}
}{%
    \newcommand{\add}[1]{#1}
    \newcommand{\del}[1]{\ignorespaces}
    \newcommand{\com}[1]{\ignorespaces}
}

\onlineid{0}

\vgtccategory{Research}

\vgtcinsertpkg

\title{\add{Transtiff: A Stylus-shaped Interface for Rendering Perceived Stiffness of Virtual Objects via Stylus Stiffness Control}}

\author{Ryoya Komatsu\thanks{e-mail: ryoya.komatsu@x-lab.team}\\ %
        \scriptsize Aoyama Gakuin University %
\and Ayumu Ogura\thanks{e-mail: ayumu.ogura@x-lab.team}\\ %
     \scriptsize Aoyama Gakuin University %
\and Shigeo Yoshida\thanks{e-mail: shigeo.yoshida@sinicx.com}\\ %
     \scriptsize OMRON SINIC X Corporation %
\and Kazutoshi Tanaka\thanks{e-mail: kazutoshi.tanaka@sinicx.com}\\ %
     \scriptsize OMRON SINIC X Corporation %
\and Yuichi Itoh\thanks{e-mail: itoh@it.aoyama.ac.jp}\\ %
     \scriptsize Aoyama Gakuin University} %

\teaser{
  \centering
  \includegraphics[width=\linewidth]{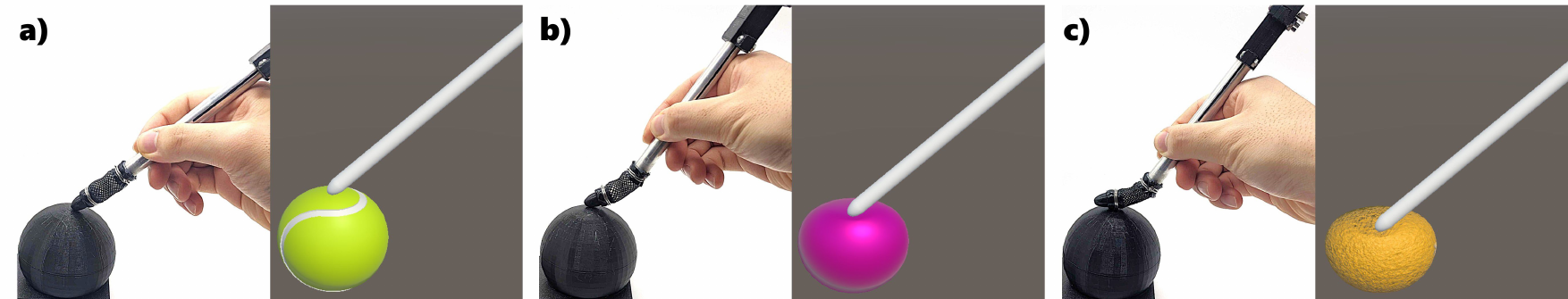}
  \caption{
  \add{Transtiff, a stylus-shaped interface, manipulates the perceived stiffness of virtual objects by controlling the stiffness of the interface, enabling the creation of the illusion that a hard object feels soft. a) Interaction with a tennis ball at maximum stiffness, b) with a plastic ball at medium stiffness, and c) with a sponge ball at minimal stiffness.}
  \del{The interface bends and the visual stimulus causes the object to be concave, allowing haptic rendering. The three states of our stiffness-changing haptic interface investigated in our user study. 
a) shows touching a tennis ball at maximum stiffness at the interface.
b) shows touching a plastic ball at half stiffness at the interface.
c) shows touching a sponge ball at minimum stiffness at the interface.}}
  \label{fig:teaser}
}

\abstract{
The replication of object stiffness is essential for enhancing haptic feedback in virtual environments. 
However, existing research has overlooked how stylus stiffness influences the perception of virtual object stiffness during tool-mediated interactions. 
To address this, we conducted a psychophysical experiment demonstrating that changing stylus stiffness combined with visual stimuli altered users' perception of virtual object stiffness. 
Based on these insights, we developed Transtiff, a stylus-shaped interface capable of on-demand stiffness control using a McKibben artificial muscle mechanism.
Unlike previous approaches, our method 
manipulates the perceived stiffness of virtual objects via the stylus by controlling the stiffness of the stylus without altering the properties of the real object being touched, creating the illusion of a hard object feeing soft.
Our user study confirmed that Transtiff effectively simulates a range of material properties, such as sponge, plastic, and tennis balls, providing haptic rendering that is closely aligned with the perceived material characteristics. 
By addressing the challenge of delivering realistic haptic feedback through tool-based interactions, Transtiff represents a significant advancement in the haptic interface design for VR applications.
} % end of abstract


\begin{document}


\firstsection{Introduction}
\label{1}

\maketitle

% VRと触覚。VRオブジェクトに触るために様々なデバイスが提案されている。その中でもスタイラスタイプのもある
With the rapid advancement of virtual reality (VR) technology, haptic feedback has become increasingly important alongside visual and auditory presentation. 
A growing body of research focuses on simulating the sensation of interacting with virtual objects via tools~\cite{perret2018touching, pupop, variable_shape, pacapa, thor, hapsticks, Drag}.
Among these, stylus-shaped interfaces~\cite{Touchx, Haptylus, ImpAct, FeelPen} have gained attention due to their ability to provide both precise control and haptic feedback during interactions. 
% 仮想現実 (VR) 技術の急速な進歩により、視覚や聴覚によるプレゼンテーションと並んで触覚フィードバックの重要性が高まっています。ますます多くの研究が、ツールを介して仮想オブジェクトとインタラクションする感覚をシミュレートすることに焦点を当てています~\cite{perret2018touching、pupop、variable_shape、pacapa、thor、hapsticks、Drag}。これらの中で、スタイラス型のインターフェース~\cite{Touchx、Haptylus、ImpAct、FeelPen}は、インタラクション中に正確な制御と触覚フィードバックの両方を提供できることから注目を集めています。

% 道具を身体化できる。道具に細工することで道具を介して触る物体の触覚知覚を変容できるかも
Humans can naturally integrate tools as extensions of their bodies, and haptic information perceived through tools is processed similarly to skin-mediated sensations~\cite{miller2019somatosensory}. 
It has been shown that when via a stylus, haptic feedback from the stylus itself influences the perception of softness and hardness of the target object~\cite{Softness_Discrimination}.
Furthermore, the stiffness of the stylus can influence the user’s haptic perception of the object, even when interacting with the same object.
%人間は道具を身体の延長として自然に統合することができ、道具を通して知覚される触覚情報は皮膚を介した感覚と同様に処理されます~\cite{miller2019somatosensory}。スタイラスを使用する場合、スタイラス自体からの触覚フィードバックが対象物の柔らかさや硬さの知覚に影響を与えることが示されています~\cite{Softness_Discrimination}。さらに、スタイラスの硬さは、同じ物体と対話している場合でも、ユーザーの物体に対する触覚知覚に影響を与える可能性があります。

% 実験してみる
In this paper, we investigate how haptic perception is influenced when objects are touched via styluses with varying levels of stiffness \add(\cref{3}). 
Although previous research has extensively explored the perception of an object’s stiffness and texture~\cite{Softness_Discrimination, klatzky1999tactile, Cavdan, Okamoto, Higashi, Cristiano, Bergmann, Xu}, the role of the stylus’s stiffness in shaping haptic perception remains underexplored. 
By investigating this, we aim to deepen our understanding of how the stiffness of a stylus influences haptic perception.
%本論文では、さまざまなレベルの硬さを持つスタイラスを使用して物体に触れたときに、触覚知覚がどのように影響を受けるかを調査します。これまでの研究では、物体の硬さと質感の知覚が広範囲に調査されてきましたが、触覚知覚の形成におけるスタイラスの硬さの役割はまだ十分に調査されていません。これを調査することで、スタイラスの硬さが触覚知覚にどのように影響するかについての理解を深めることを目指します。
% 錯覚の可能性を検証
We also explore the possibility of creating an illusion where touching a hard object with a soft stylus generates the perception of a soft object. 
This approach could enable the reproduction of haptic sensations for both hard and soft virtual objects by simply controlling the stiffness of the stylus.
%また、柔らかいスタイラスで硬い物体に触れると柔らかい物体の知覚が生まれるという錯覚を作り出す可能性も探っています。このアプローチにより、スタイラスの硬さを調整するだけで、硬い仮想物体と柔らかい仮想物体の両方の触覚を再現できるようになります。

% 結果をもとにデバイスを設計
Based on these findings, we propose a stylus-shaped interface with joints capable of controlling its stiffness (\cref{fig:teaser}) \add{(\cref{4})}. 
Our interface leverages an artificial muscle mechanism to control stiffness by controlling the pressure within a flexible tube via a piston system. 
We evaluated the stiffness control capabilities of the interface, focusing on its ability to reproduce the sensation of touching objects with varying stiffness levels and to its effectiveness of the interface in enhancing haptic experience in VR \add{(\cref{5})}.
%これらの調査結果に基づいて、剛性を動的に変更できるジョイントを備えたスタイラス型のインターフェースを提案します。私たちのインターフェースは、人工筋肉メカニズムを利用して、ピストンシステムを介して柔軟なチューブ内の圧力を制御することで剛性を調整します。インターフェースの剛性調整機能を評価する実験を行いました。これらの実験は、さまざまな剛性レベルのオブジェクトに触れる感覚を再現し、VR での触覚フィードバックを強化するインターフェースの有効性を評価するために設計されました。

The main contributions of this paper are the following:
\begin{enumerate}[label=\textbf{(\arabic*)}]
  \item We demonstrated through our psychophysical experiment that touching a hard object with visual stimuli of object softness and a soft stylus may create an illusion that the hard object felt soft.
  \item  We implemented a stylus-shaped interface with stiffness control using a McKibben-type artificial muscle mechanism to apply the above finding for haptic rendering of various object stiffnesses.
  \item We confirmed that the proposed interface can alter the perceived stiffness of virtual objects through the user study.
\end{enumerate}

\add{
The experiments described in this paper were conducted with the approval of the Ethical Review Committee of Aoyama Gakuin University (approval number: H22-008).
}
% 我々は心理物理学的実験を通じて、物体の柔らかさの視覚刺激と柔らかいスタイラスで硬い物体に触れると、その物体が柔らかく知覚されるという錯覚が生じる可能性があることを実証した。
%我々は、マッキベン型人工筋肉機構を使用して動的剛性制御を備えたスタイラス型インターフェースを導入し、実装し、上記の発見をさまざまな物体の剛性の触覚レンダリングに適用した。
%ユーザスタディを通じて、提案されたインターフェースが物体の知覚される剛性を変更できることを確認した。

\section{Related Work}
\label{2}

In this section, we review related works on stylus-shaped interfaces for haptic feedback and discuss haptic perception via the stylus.
We also briefly overview the current technologies of soft actuators and mechanisms to contextualize our device design.
%このセクションでは、まず触覚フィードバック用のスタイラス型インターフェースに関する関連研究をレビューし、次にスタイラスによる触覚知覚について説明します。また、デバイス設計の位置づけを明確にするために、ソフトアクチュエータの現在の技術についても簡単に説明します。

\subsection{Stylus-Shaped Interface for Haptic Feedback}
\label{2_1}

Several haptic systems have been proposed to generate sensations of touching virtual objects~\cite{MultimodalHapticDisplay}, one of which involves the use of stylus-shaped interfaces to deliver haptic feedback. 
A notable example is the Touch X~\cite{Touchx} by 3D Systems, which provides force feedback via a stylus in an indirect manner when users interact with objects displayed on a PC screen. 
In contrast, Withana \etal\ introduced ImpAct,~\cite{ImpAct} an interface that can change its length when pushed directly against a surface to provide kinesthetic haptic sensation to the hand.
Haptylus~\cite{Haptylus} employs similar mechanisms with ImpAct that can alter the length of the stylus, but it incorporates vibrations via a voice coil motor to simulate the texture of virtual objects. 
FeelPen~\cite{FeelPen} offers a wide range of sensations such as viscosity, roughness, friction, and temperature, combining various modalities to create a more immersive experience.
%仮想空間で感覚を生成するために、いくつかの触覚提示方法が提案されている。顕著な例として、3D Systems社の ``Touch X''があり、ユーザーがPC画面上に表示されたオブジェクトとインタラクションする際に、スタイラスを通してフォースフィードバックを提供する。Withanaらは、触覚フィードバックを提供するために、表面に押し当てたときにスタイラス先端の長さを調整するデバイス、 ``ImpAct''を発表した。同様に、長坂らの「Haptylus」は、スタイラス先端の長さを変え、ボイスコイルモーターを介して振動を取り入れ、仮想物体の質感をシミュレートする。Benceらの ``FeelPen''は、粘性、ざらつき、摩擦、温度など幅広い感覚を提供し、さまざまなモダリティを組み合わせることで、より没入感のある体験を生み出す。
\del{
These interfaces use internal power units to deliver haptic feedback directly to a user’s hand, thereby reproducing a virtual sense of contact. 
}
%これらのデバイスは、内蔵のパワーユニットを使ってユーザーの手に直接触覚フィードバックを与え、バーチャルな接触感覚を再現します。
% \add{
% These interfaces rely on internal power units to directly transmit haptic feedback to the user’s hand, thereby effectively recreating the sensation of contact with virtual objects.
% }
%これらのデバイスは、内部の電源ユニットを使用して触覚フィードバックをユーザーの手に直接伝え、仮想オブジェクトとの接触感覚を効果的に再現します。
\add{
Lee \etal\ proposed a Haptic Pen~\cite{Hapticpen} that improves user interaction with touch screens by delivering localized vibrations and force cues. 
In a similar vein, Fellion \etal\ introduced FlexStylus~\cite{FlexStylus}, a stylus interface that incorporates bend input as an additional interaction modality, thereby broadening the potential applications of pen-based systems.
}
%Leeらは、局所的な振動と力の合図を提供することでタッチスクリーンとのユーザーインタラクションを改善するHaptic Pen [23]を提案しました。同様に、Fellionらは、追加のインタラクション様式として曲げ入力を組み込んだスタイラスインターフェースであるFlexStylus [9]を発表し、ペンベースシステムの潜在的な用途を広げました。
\add{
Through these developments, stylus-shaped interfaces have demonstrated their ability to enhance touch-based interactions with screens by offering diverse haptic feedback mechanisms and interaction capabilities that are tailored to various scenarios.
}
%これらの開発を通じて、スタイラス型インターフェースは、さまざまなシナリオに合わせて調整された多様な触覚フィードバックメカニズムとインタラクション機能を提供することで、画面とのタッチベースのインタラクションを強化する能力を実証しました。

\del{
In contrast, some interfaces create varied haptic experiences by altering the characteristics of the stylus itself. 
}
%しかし、これらの実装では、スタイラスの硬さは一定のままである。
\add{
In a related approach, some studies focused on altering the physical characteristics of the stylus to provide varied haptic experience. 
}
%関連するアプローチとして、いくつかの研究では、スタイラスの物理的特性を変更してさまざまな触覚体験を提供することに焦点を当てました。
Liu \etal\ proposed the FlexStroke~\cite{FlexStroke}, a pen interface that uses a jamming transition mechanism to replicate the haptic sensations of a Chinese brush, an oil brush, and crayon by controlling the air pressure at the tip of the interface, thereby providing variable stiffness. 
%Liuらは、「FlexStroke」というペンインターフェースを提案しました。このインターフェースは、デバイスの先端の空気圧を制御することによって硬さを変化させ、中国筆、油筆、クレヨンの触感を再現するジャミング遷移機構を使用しています。
\add{
Kara presented the VnStylus~\cite{VnStylus}, a haptic stylus that integrates variable tip compliance to enhance haptic rendering of virtual environments.
}
%Karaは、可変先端コンプライアンスを統合して仮想環境の触覚レンダリングを強化する触覚スタイラスである VnStylusを発表しました。
Moreover, we have been proposing a stylus-shaped haptic interface, Transtiff~\cite{ogura_uist, ogura_siggraph}, with a mechanism that incorporates an artificial muscle.
The interface features a joint that controls its stiffness, thus altering the force required for bending.
%また，小倉らは，人工筋肉機構を備えた棒状触覚デバイス ``Transtiff''を提案している．棒の中継部に硬軟が変化する関節を作成し，曲げるのに必要な力を変化させている．
Although these studies successfully control haptic feedback by varying the stylus stiffness, they did not investigate how varying stylus stiffness affects the perception of objects touched via the stylus.
%これらの研究ではスタイラス自体の硬さを変えることで触覚フィードバックを制御する方法が提案されている．しかし，主にスタイラス型インターフェースの技術的な実装に焦点を当てており、剛性の異なるスタイラスを介した物体の知覚を十分に調査していません。
Specifically, it remains unclear whether users perceive changes in stiffness as an attribute of the stylus or an object touched by the stylus.  

Therefore, we believe that clarifying how users perceive stiffness when interacting with objects via styluses of different stiffness levels will deepen our understanding of haptic presentation via stylus-shaped interfaces and unlock new possibilities for application in VR.
%したがって、異なる硬さのスタイラスを通してオブジェクトとインタラクションする際に、ユーザーがどのように硬さを知覚するかを明らかにすることは、スタイラス型デバイスを使用した触覚提示の理解を深め、VR環境でのアプリケーションの新たな可能性を引き出すと信じています。

\subsection{Haptic Perception via Stylus}
\label{2_2}

Regarding haptic perception via a stylus, LaMotte \etal\ demonstrated that stiffness discrimination is possible when touching a flexible object via a stylus, with discrimination accuracy particularly enhanced by making a tapping motion~\cite{Softness_Discrimination}. 
Similarly, Klatzky \etal\ investigated roughness perception when touching a rough surface through a rigid tool (\eg, a stylus-shaped probe) and found that roughness discrimination was achievable through a rigid body, albeit with less accuracy than when using a fingertip~\cite{klatzky1999tactile}. 
Chung \etal\ reported comparable results with a rubber-tipped stylus. 
Kato \etal\ further showed that their proposed chopstick-like device for virtual weight presentation resulted in a lower weight discrimination rate compared to the perception of real objects~\cite{Kato}.
%棒を介した触覚知覚についてLaMotteらは，スタイラスを介して柔軟な物体に触れる際の硬軟識別が可能であることを示し，「叩く」動作をすることで硬軟の識別率が特に増加することを明らかにしている \cite{Softness_Discrimination}．Klatzkyらは，剛体（棒状のプローブ等）を通して粗面に触れる際の粗さ知覚を調査しており，指腹より精度は低いが，剛体を通しても粗さの識別が可能なことを明らかにしている \cite{klatzky1999tactile}．Chungらは，先端がゴム製のスタイラスと弾性表面のディスプレイがタップ動作に最適であることを示している \cite{CHUNG201512}．Katoらは，提案する箸型デバイスの仮想重量感提示によって，実物体の重量知覚よりも重量の識別率が低下する錯覚が生じると報告している \cite{Kato}．

Numerous studies have examined haptic perception using force-sensing interfaces similar to the one such as Touch X~\cite{Touchx}.
Heather \etal\ reported that when recreating a surface in a virtual environment using a 3D haptic interface, the perceived intensity of slipperiness, stiffness, and roughness could be varied by presenting transient friction and contact stimuli along with surface textures~\cite{Importance}.
%\ref{2-1}節で前述した力覚デバイス \cite{Touchx}と同機能を持つ装置を用いて知覚調査をしている事例も多い．Heatherらは，3次元触覚デバイスを用いてバーチャル空間の表面を再現する際，摩擦・接触刺激の過渡現象・表面テクスチャを提示することで，ユーザが感じる滑りやすさ・硬さ・粗さの強度を変化させられると報告している \cite{Importance}．
\add{
These findings highlight the significance of combining multiple sensory modalities in creating realistic virtual environments.
}
%これらの研究結果は、リアルな仮想環境を作成するために複数の感覚様式を組み合わせることの重要性を強調しています。

\add{
While these studies focused on the mechanical properties of objects and tools, the interaction between visual and haptic cues also plays a key role in shaping users' perception of object properties even when physical feedback is constant. A notable example is pseudo-touch, in which visual cues affect haptic sensations.  
%これらの研究は物体や道具の機械的特性に焦点を当てていますが、視覚的手がかりと触覚的手がかりの相互作用も、物理的なフィードバックが一定である場合でも、物体の特性に対するユーザーの認識を形成する上で重要な役割を果たします。注目すべき例は、視覚的手がかりが触覚感覚に影響を与える疑似タッチです。
Although pseudo-touch has been extensively studied in fingertip interactions~\cite{Pseudo-Stiffness}, its role in stylus-based systems remains underexplored. Our findings bridge this gap by showing how compliant styluses leverage visual cues to enhance haptic interaction fidelity, particularly through stiffness control.  
%疑似タッチは指先インタラクションで広く研究されてきましたが~\cite{疑似剛性}、スタイラスベースのシステムでの役割はまだ十分に調査されていません。私たちの研究結果は、柔軟なスタイラスが視覚的手がかりを活用して、特に動的剛性制御を通じて触覚インタラクションの忠実度を高める方法を示すことで、このギャップを埋めています。
}

\add{
Prior studies have largely overlooked the influence of stylus stiffness on haptic perception. This study addresses this gap by examining how varying stylus stiffness affects perception, laying the groundwork for designing haptic interfaces that use stylus stiffness as a controllable parameter to enhance virtual interaction.
%これまでの研究では、スタイラスの剛性が触覚知覚に与える影響はほとんど見過ごされてきました。この研究では、スタイラスの硬さの変化が知覚にどのように影響するかを調べることでこのギャップに対処し、仮想インタラクションを強化するためにスタイラスの硬さを制御可能なパラメータとして使用する触覚インターフェースを設計するための基礎を築きます。
}

\del{
Although various studies have investigated haptic perception through styluses, to our knowledge, none have explored how the stiffness of the stylus itself influences haptic perception. 
In contrast, this study investigates the effects on haptic perception when an object is touched with styluses of different stiffness and applies these findings to the design of haptic interfaces.
}
%このように，棒を介した触覚知覚について複数の調査がされているものの，いずれも棒自体の硬軟が異なる場合の触覚知覚については言及されていない．対して，本研究では，硬軟の異なる棒を用いて物体に触れたときの触覚知覚に与える影響を調査して，触覚インタフェースに応用する．

\subsection{Soft Actuators \add{and Mechanisms}}\label{2_3}

Research on soft actuators \add{and mechanisms} for joints capable of flexibly altering their stiffness has progressed, contributing to advancements in soft robotics technology \add{and relevant fields}.
The commonly used actuation method is controlling the expansion, contraction, and flexion of actuators through pneumatic systems, which have been applied in robotic hands~\cite{abondance2020dexterous} and power-assist devices~\cite{yap2015soft}. 
Among these, pneumatic actuators, specifically ``McKibben-type artificial muscles ''~\cite{tondu2012modelling, zhang2017modeling, Schulte1961TheCO, meller2016improving}, are frequently used.
%ソフトロボティクス技術の発展に伴い，柔軟な剛性変化が可能な関節の研究が進んでいる．特に，空気圧によってアクチュエータの伸縮や屈曲を制御する手法はよく用いられており，ロボットハンドや \cite{abondance2020dexterous}，パワーアシストデバイスなどに利用されている \cite{yap2015soft}．これらの空気圧アクチュエータは，2層のチューブ内で流体を加圧することにより軸方向へ収縮を起こす ``マッキベン型人工筋肉'' \cite{tondu2012modelling, zhang2017modeling, Schulte1961TheCO, meller2016improving}が提案されている． 
Devices utilizing McKibben-type artificial muscles have been applied in various applications, such as tele-surgery~\cite{ForcepsManipulator}, shape-changing interfaces~\cite{MckkibenInteraction}, and haptic gloves~\cite{Use_of_McKibben_Muscle}.

Others have proposed the combination of pneumatic soft actuators with jamming transitions. 
Wall \etal\ introduced a soft actuator using layer jamming, a technique that induces a jamming transition in overlapping sheets, to achieve significant stiffness changes~\cite{layerjaming}. 
%他にも、空気圧式ソフトアクチュエータとジャミング遷移の組み合わせを提案した研究者もいる。Wallらは、重なり合ったシートにジャミング遷移を誘発する技術であるレイヤージャミングを使用して、大幅な剛性の変化を実現するソフトアクチュエータを導入した。
This actuator allows for more efficient modulation of stiffness and softness than conventional approaches. 
However, it limits the actuators' bending direction to perpendicular to the sheet layers, making it difficult to bend parallel to them.
%このアクチュエータは、従来のアプローチよりも効率的に剛性と柔らかさを調整できます。ただし、アクチュエータの曲げ方向はシート層に対して垂直に制限されるため、シート層と平行に曲げることは困難です。
To address this, Yang \etal\ developed a system that can 3D print variable stiffness jamming interfaces in arbitrary shapes~\cite{MultiJam}. 
By configuring the arrangement of membranes and beads, this system can print haptic interfaces and shape-changing controllers that function like soft actuators.
However, a common challenge for jamming-based mechanisms is the need for a large external air compressor.
%これに対処するため、Yang らは、任意の形状の可変剛性ジャミング インターフェースを 3D プリントできるシステムを開発しました ~\cite{MultiJam}。このシステムは、膜とビーズの配置を構成することで、ソフト アクチュエータのように機能する触覚インターフェースと形状変更コントローラーをプリントできます。ただし、ジャミング ベースのメカニズムに共通する課題は、大型の外部エアコンプレッサーが必要になることです。

In response to the challenge of requiring external compressor, many soft actuators utilizing the characteristics of functional materials have been developed. 
For example, low-melting-point alloy (LMPA)~\cite{hao2018eutectic} and shape-memory polymers (SMP)~\cite{zhang2019fast} have been used as actuators, where temperature control to adjust the stiffness variation.
Magnetorheological (MR) fluids, which change the stiffness based on the strength of the applied magnetic field~\cite{de2011magnetorheological}, is also widely used.  
Ohba \etal\ proposed a linear motion joint that enables soft motion by adjusting the holding force of a spring through the magnetic field control of the MR fluid surrounding a compression spring~\cite{ohba2012elastic}.
Additionally, electrorheological (ER) fluids, which change liquid viscosity by an electric field, are used in soft robots, often in contrast to MR fluids~\cite{electrorheological}. 
However, these functional materials can be difficult to obtain and manage, limiting their ease of use.
%機能性材料の持つ特徴を利用するソフトアクチュエータも多く開発されている．Haoらは，低融点合金（LMPA）を埋め込み，ヒータで液体--固体の状態変化を起こすことで，従来よりも剛性の変化幅が大きいアクチュエータを提案している \cite{hao2018eutectic}．Zhangらは，形状記憶ポリマー(SMP)をアクチュエータに埋め込むことにより，温度制御でポリマーの剛性を変化させ，耐荷重を向上させている \cite{zhang2019fast}．また，磁界の強度によって剛性が変化する流体であるMR流体 \cite{de2011magnetorheological}もよく利用されている．大場らは，圧縮ばねの周りに満たしたMR流体にかける磁界を制御することで，ばねの保持力を変化させ，やわらかい動作を実現する直動型関節を提案している \cite{ohba2012elastic}．一方で，電場を制御することで液体の粘性が変化する電気粘性流体（ER流体）も，ソフトロボットにおける構成要素として用いられており，MR流体との使い分けが進んでいる \cite{electrorheological}．しかし，これらの機能性材料を用いた手法に関しては，入手や管理が難しく，容易に使用できないという難点がある．

\add{
As another stiffness-changing mechanism, Ryu \etal\ proposed ElaStick~\cite{ElaStick}, a handheld device that dynamically modulates the stiffness of four elastic tendons for 2-DoF haptic feedback, simulating flexible objects, but its size limits integration into stylus-shaped interfaces.
}
%剛性を変える別のメカニズムとして、Ryuらは、柔軟なオブジェクトをシミュレートする 2-DoF 触覚フィードバックのために 4 つの弾性腱の剛性を動的に調整するハンドヘルド デバイスである ElaStick を提案しましたが、そのサイズによりスタイラス形状のインターフェイスへの統合が制限されます。

While methods like layer jamming and functional materials have limitations, we believe that artificial muscles, controlled by simple fluid manipulation, offer a more suitable and straightforward solution for stiffness control in stylus-shaped interfaces.
Consequently, we employed McKibben-type artificial muscles to develop an interface capable of controlling stiffness.
\add{The details of the proposed interface are described in \cref{4}.}
%以上のように，柔軟な剛性変化関節を実現しようとする研究は多い．レイヤージャミングを用いた手法では，関節が曲がる方向が限定されてしまい，触覚デバイスとして制限が生じてしまう．また，機能性材料を用いた手法に関しては，入手や管理が難しいという難点がある．これらの点から，単純な流体操作のみで実現できる人工筋肉の構造が，触覚デバイスへの応用に適していると考えられる．

\section{Psychophysical Experiment with Stylus and Objects of Various Stiffnesses}
\label{3}
\add{
The aim of this work is to deepen the understanding of how stylus stiffness influences haptic perception and explore its potential for creating haptic illusions that a hard object felt soft when interacting with a hard object.  
Specifically, this experiment examines the effects of varying levels of stylus stiffness on the perception of objects during touch.
%この研究の目的は、スタイラスの硬さが触覚知覚にどのように影響するかについての理解を深め、触覚錯覚を生み出す可能性を探ることです。具体的には、この研究では、さまざまなレベルのスタイラスの硬さが、タッチ時の物体の知覚に与える影響を調べました。
}
\del{
We conducted an experiment where users touched an object with a stylus of varying stiffness, combined with a visual stimulus, to investigate how the perception of the object is affected by the visual stimulus and the stiffness of the stylus.
}
%本章では，視覚刺激とスタイラスの剛性によって物体の知覚にどのような影響があるのか調査するために，視覚刺激と合わせて様々な剛性のスタイラスで物体に触れる実験を行う．

\begin{figure}[t]
    \centering
    \includegraphics[width=80mm]{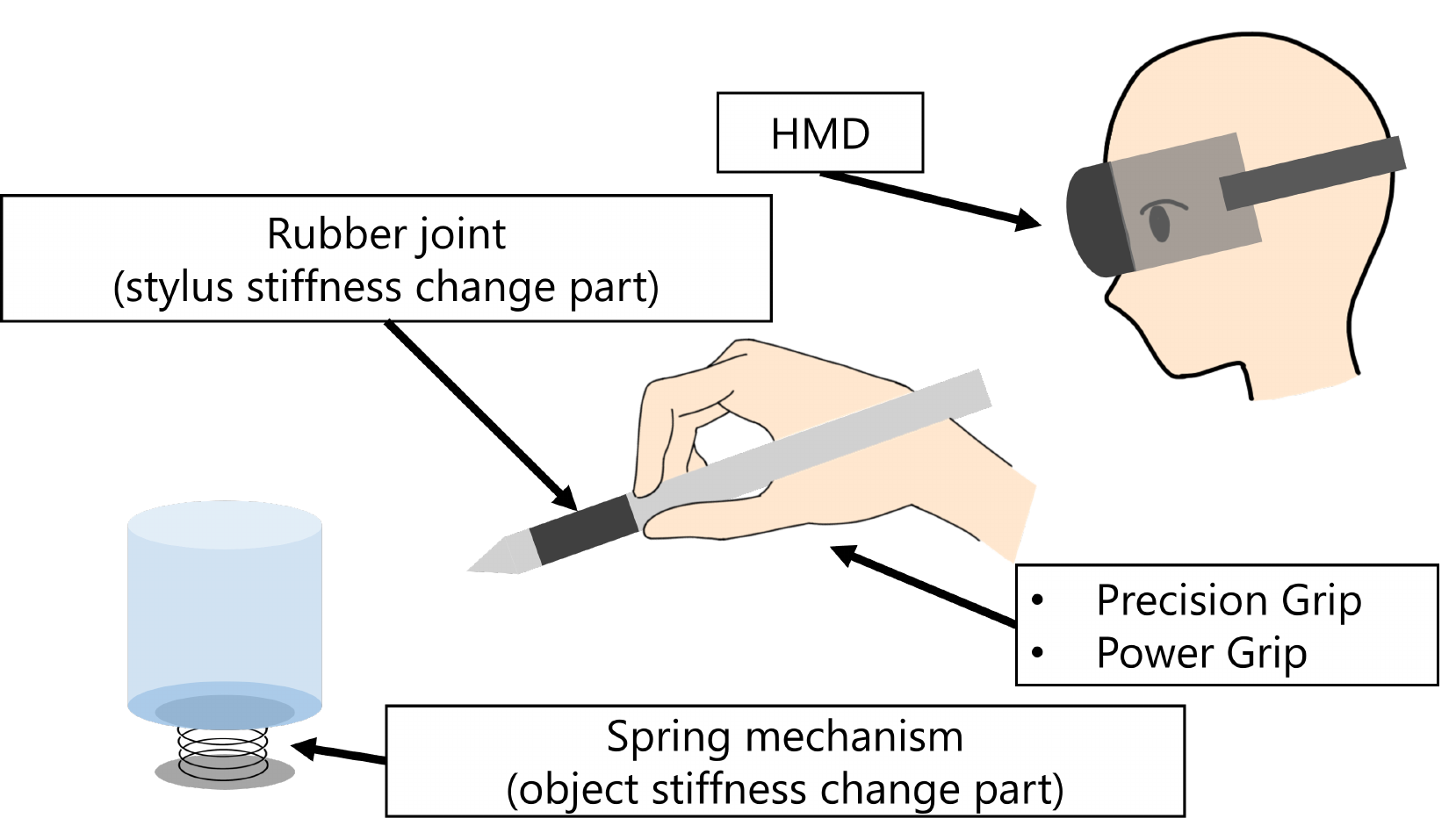} 
    %\vspace{-1cm}
    \caption{Experimental overview. \add{The participants use the HMD to receive visual stimuli showing the deformation of the stylus or object and manipulate a stylus with a rubber joint to explore objects of different stiffness (a spring-loaded cylindrical mechanism).}}
    %参加者は HMD を装着し、スタイラスまたは物体の変形を示す視覚刺激を受け取りました。参加者はゴム製のジョイントを備えたスタイラスを操作して、さまざまな硬さの物体を探査しました (バネ仕掛けの円筒形機構)。
    \label{fig:gaiyou}
\end{figure}

We examined whether users could discriminate which was softer -- the stylus or the object -- when they touched the object via the stylus while showing a visual stimulus of either the stylus or the object was deformed.
%スタイラスと物体のどちらかが変形する視覚刺激を提示して，スタイラスを用いて物体に触れる際に，スタイラスと物体のどちらが軟らかいかをユーザが判別できるかを私達は調査した．
\add{
Furthermore, we investigated the frequency with which participants experienced the illusion that a hard cylinder felt soft when via a soft stylus in combination with specific visual stimuli.
}
%さらに、柔らかいスタイラスを特定の視覚刺激と組み合わせて使用したときに、硬い円筒が柔らかく感じられるという錯覚を参加者が経験する頻度を調査しました。
We also investigated whether the presentation of visual stimuli induces cross-modal perception~\cite{moody2009beyond}, altering the perceived stiffness of styluses and objects.
%また，視覚刺激の提示によってクロスモーダル知覚 \cite{moody2009beyond}が生起され，棒と物体に対する硬さ知覚が変化するのかについても調査する．
\cref{fig:gaiyou} shows an overview of the experiments.  
Participants grasped a stylus with a flexible rubber joint and used it to touch an object with varying stiffness (a cylinder equipped with a spring mechanism). 
%図 \ref{fig:gaiyou} は実験の概要を示しています。参加者は HMD を装着し、スタイラスまたはオブジェクトの変形を示す視覚刺激を受け取りました。参加者は、ゴム製のジョイントを備えたスタイラス (バネ仕掛けの円筒形機構) を操作して、さまざまな剛性を持つオブジェクトを探りました。このセットアップの目的は、視覚と触覚の合致により、参加者がスタイラスまたはオブジェクトをより柔らかく感じるという触覚錯覚を誘発することでした。この実験は、動的剛性制御がこの錯覚を効果的にサポートするかどうかを評価するために設計されました。

\subsection{Experimental Condition}
\label{3_1}

\begin{figure}[t]
    \centering
    \includegraphics[width=80mm]{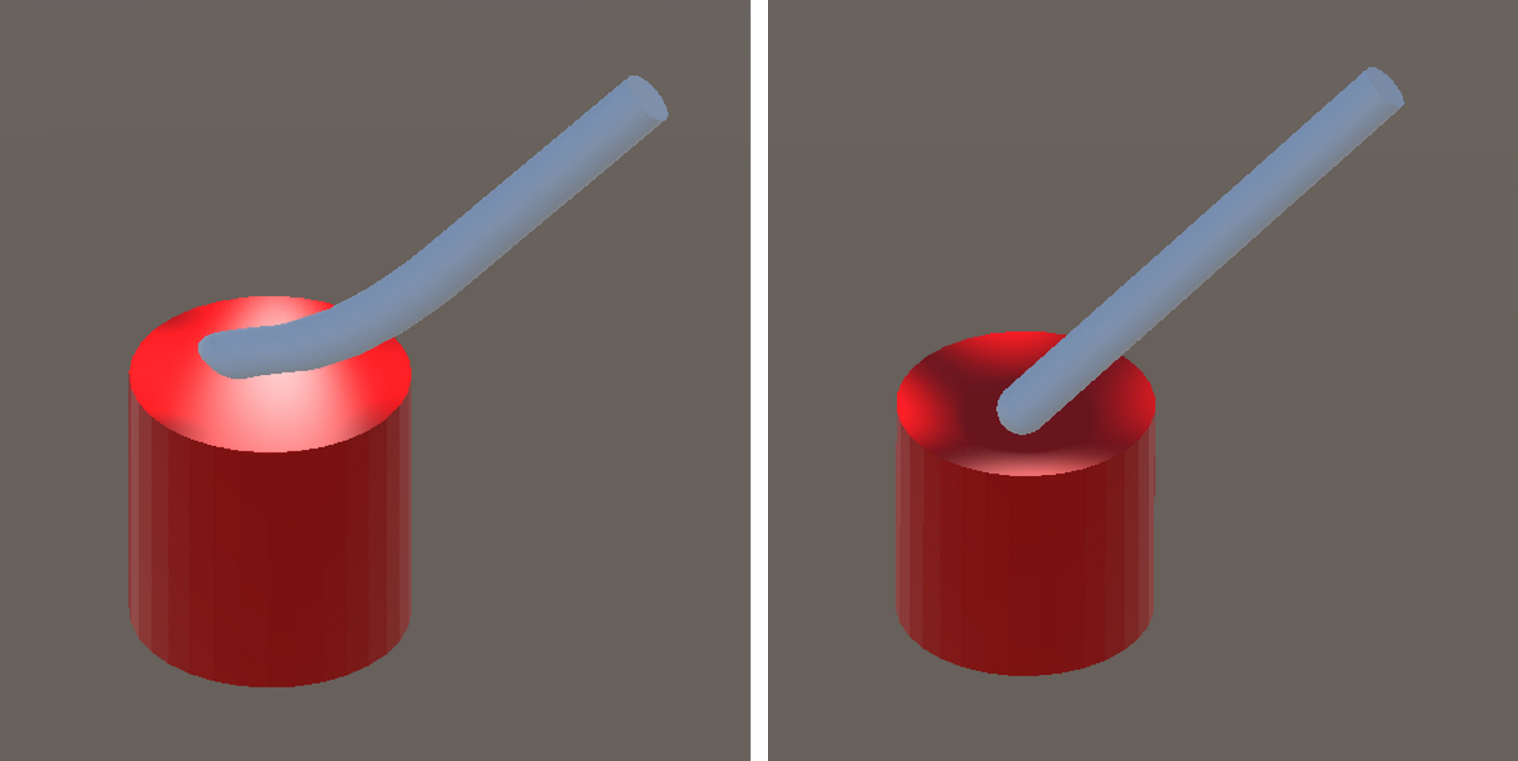} 
    %\vspace{-1cm}
    \caption{Stylus/Cylinder deformation. \add{The visual stimuli consisted of a stylus appearing soft on the left and a cylinder appearing soft on the right, both designed to visually convey stiffness. }}
    %視覚刺激は、左側の柔らかく見えるスタイラスと右側の柔らかく見える円筒で構成され、どちらも硬さを視覚的に伝えるように設計されています。
    \label{fig:bendable}
    %\vspace{-5mm}
\end{figure}

In this experiment, seven stylus and five cylindrical contact objects with varying levels of stiffness were prepared. Furthermore, two conditions were presented as visual stimuli through an HMD (HTC VIVE Pro2): the stylus was deformed or the cylinder was deformed when the stylus made contact with it. 
%本実験では，硬さの異なる7条件の棒と，5条件の円柱を用意した．さらに，棒を用いて円柱に触れる際に，棒が変形，もしくは，円柱が変形する2条件を，HMD（HTC VIVE Pro2）を通して視覚刺激として提示した．
\add{
To investigate the relationship between stylus stiffness and haptic perception, these conditions were designed to enable a controlled range of stylus stiffness that induced visually discernible macroscopic deformation, effectively creating a haptic illusion that a hard object felt soft during interactions with a hard object via a soft stylus.
}
%スタイラスの硬さと触覚知覚の関係を調査するために、これらの条件は、視覚的に識別可能なマクロ変形を誘発する制御された範囲のスタイラスの硬さを可能にし、柔らかいスタイラスを介して硬い物体と相互作用する際に、柔らかさの触覚錯覚を効果的に作り出すように設計されました。
\cref{fig:bendable} illustrates the deformation of the stylus and the cylinder. 
%図\ref{fig:bendable}に棒・円柱の変形を示す．視覚刺激は、左側の柔らかく見えるスタイラスと右側の柔らかく見える円筒で構成されており、どちらも硬さを視覚的に伝えるように設計されています。これらのオブジェクトは、スタイラスとの相互作用に応じて動的に変形し、動的硬さレンダリングにおける触覚知覚を調査できます。
\add{
To present visual stimuli, we developed a virtual environment using Unity to simulate a stylus interacting with a cylindrical object. We controlled animations to replicate varying levels of simulated stiffness using VertExmotion\footnote{\url{https://assetstore.unity.com/packages/tools/animation/vertexmotion-23930?locale=ja-JP}} asset, which implements a shader-based soft-body system. This feature allowed us to investigate how changes in apparent stiffness influenced user interaction and perception, thereby providing a foundation for evaluating the accuracy of haptic feedback models.
}
%視覚刺激を提示するために、Unity を使用して仮想環境を開発し、円筒形のオブジェクトと相互作用するスタイラスをシミュレートしました。シェーダーベースのソフトボディシステムを実装する VertExmotion アセットを使用して、アニメーションを動的に調整し、シミュレートされたさまざまなレベルの剛性を再現しました。この機能により、見かけの剛性の変化がユーザーの操作と知覚にどのように影響するかを調査することができ、触覚フィードバック モデルの精度を評価するための基礎ができました。
In the case of stylus deformation, the stylus bends when pressed against the cylinder in virtual space, providing a visual stimulus for the soft stylus. 
In the case of cylinder deformation, the stylus presses against the cylinder in the virtual space, causing the cylinder to concavity and providing the visual stimulus of a soft cylinder.
%棒が変形する場合では，バーチャル空間で棒が円柱に接触・加圧されることで棒が曲がり，軟らかい棒という視覚刺激を提示する．円柱が変形する場合では，バーチャル空間で棒が円柱に接触・加圧されることで円柱が凹み，軟らかい円柱という視覚刺激を提示する．
\begin{figure}[t]
    \centering
    \includegraphics[width=80mm]{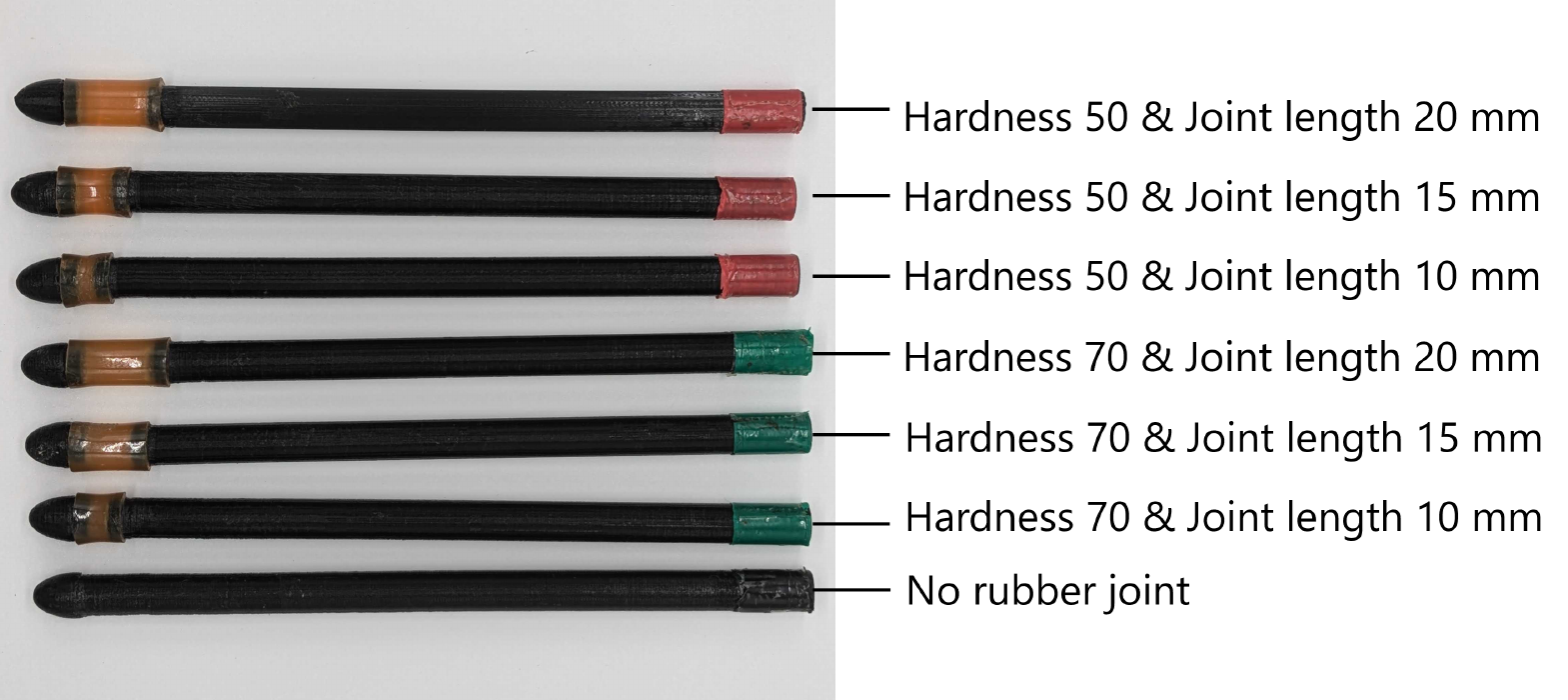} 
    %\vspace{-1cm}
    \caption{\add{Variations of the stylus used in the experiment, showing different combinations of hardness levels (50 or 70) and joint lengths (10 mm, 15 mm, or 20 mm), along with a condition without a rubber joint.}}
    \label{fig:stylus_c}
\end{figure}
\begin{figure}[t]
    \centering
    \includegraphics[width=80mm]{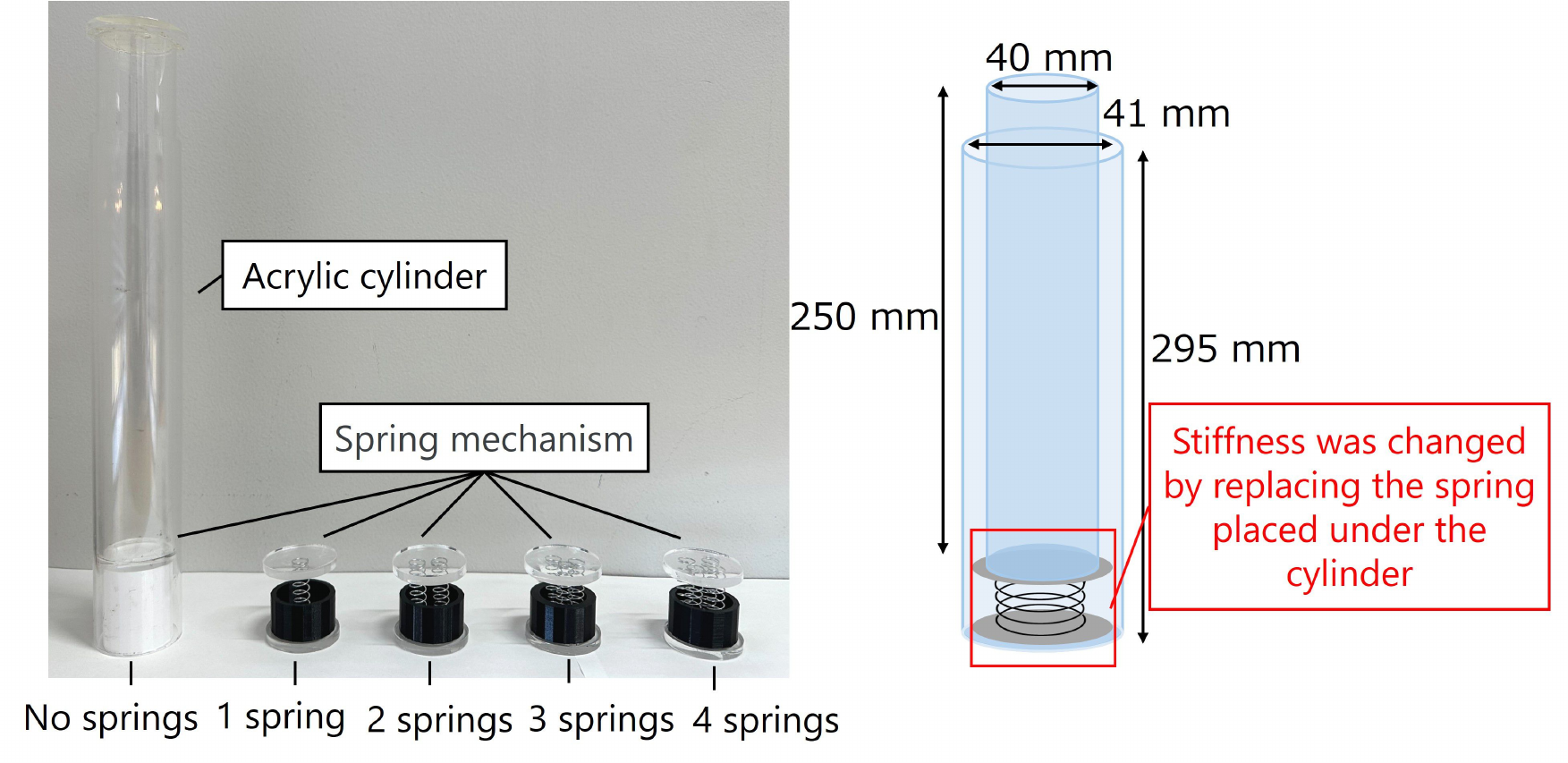} 
    %\vspace{-1cm}
    \caption{\add{Cylinder spring conditions and mechanism. The left side illustrates the five predefined stiffness levels of the cylinder's spring mechanism (spring constant: \SI{0.5}{N/mm}), while the right side depicts the cylinder size and stiffness control mechanism.}}
    %左側には、シリンダーのスプリング機構の 5 つの定義済み剛性レベル (スプリング定数: \SI{0.5}{N/mm}) が示され、右側には、機構の動的剛性調整プロセスの説明が示されています。
    \label{fig:cylinder_c}
\end{figure}
% \begin{figure}[t]
%     \centering
%     \includegraphics[width = 80mm]{figures/cp3/spring.pdf}
%     %\vspace*{-0.5cm}
%     \caption{\del{Stylus and Cylinder used in the experiment (Left. Each condition of the stylus, Right. Each condition of the cylinder)}}
%     \label{fig:spring}
% \end{figure}

A rubber joint was inserted \SI{10}{mm} from the stylus tip. This length was determined based on a preliminary experiment measuring the distance between the tip of the stylus and the user's fingertip when a typical pen was grasped spontaneously. 
The joint length was set to three values: \SI{20}{mm}, \SI{15}{mm}, and \SI{10}{mm}. A longer joint length would impair the functionality of the stylus, whereas a shorter length would prevent sufficient bending of the stylus. The stiffness of the rubber joints was set to two conditions: ``Hardness 50'' and ``Hardness 70,'' using the softest commercially available urethane round rods and those that were not excessively stiff. The higher the hardness value, the stiffer is the material. 
%実験で使用する棒は，先端から\SI{10}{mm}の部分にゴム関節を挿入している．これは，一般的なペンを自然に把持した時，先端から指までの距離を測定する予備実験をもとに，その平均値と，棒の持ち手から棒先端までが等しくなる長さになることを意図して設計している．関節長は，長すぎると棒としての機能が損なわれ，短すぎると曲がらないため，「\SI{20}{mm}」「\SI{15}{mm}」「\SI{10}{mm}」の3条件に設定した．また，ゴム関節の硬度は，市販のゴム棒（ウレタン丸棒）で最も軟らかい硬度のものと，硬すぎない硬度のものを使用するため，「硬度50」「硬度70」の2条件に設定した．硬度は数値が大きいほど硬い材料を意味する．
To the above combination, a stylus with ``no rubber joint'' is added. 
\add{\cref{fig:stylus_c} shows the seven conditions for the stylus.}
\del{
The left side of \cref{fig:spring} shows the seven conditions for the stylus from top to bottom, as shown below.
}
%上記の組み合わせに「ゴム関節なし」のスタイラスを加える．図では以下に示すように，上からの順番でスタイラスの7条件を示す．
% 箇条書き
\del{
\begin{quote}
 \begin{itemize}
  \setlength{\parskip}{0mm} % 段落間
  \setlength{\leftskip}{5mm}
  \del{
  \item Hardness 50 \& Joint length \SI{20}{mm}
  \item Hardness 50 \& Joint length \SI{15}{mm}
  \item Hardness 50 \& Joint length \SI{10}{mm}
  \item Hardness 70 \& Joint length \SI{20}{mm}
  \item Hardness 70 \& Joint length \SI{15}{mm}
  \item Hardness 70 \& Joint length \SI{10}{mm}
  \item no rubber joint
  }
 \end{itemize}
\end{quote}
}

Rubber joints allow for more significant bending with longer joint lengths or lower hardness, enabling a wide range of stiffness levels to be tested under these seven conditions. The diameter of the stylus was set to \SI{8.9}{mm}, based on the diameter of a standard stylus, specifically that of the ``Apple Pencil''~\cite{ApplePencil}, which has been commonly used as a typical stylus pen in recent years. All parts, except the rubber joints, were fabricated using a 3D printer.
%ゴム関節は関節長が長い，または硬度が低いほど曲がりやすくなる傾向を示すため，上記の7条件で広範囲の硬さを再現することができる．また，今回は棒の径が8.9 mmになるように設計した．これは，普遍的な棒状物体の径を採用することを意図し，近年代表的なスタイラスペンとして利用される ``Apple Pencil'' \cite{ApplePencil}の径を参考にしている．ゴム関節以外の部品はすべて3Dプリンタで作成した．

The cylinder mechanism was constructed by combining the two types of cylinders and a spring mechanism. 
\add{
An outer cylinder with a height of \SI{295}{mm} and an inner diameter of \SI{41}{mm} was placed around an inner cylinder with a height of \SI{250}{mm} and an outer diameter of \SI{40}{mm}. 
}
%高さ 295 mm、内径 41 mm の外筒を、高さ 250 mm、外径 40 mm の内筒の周囲に配置しました。
\del{
An outer cylinder with a height of 295 mm and an inner diameter of 41 mm was placed around an inner cylinder with a height of 250 mm and an inner diameter of 36 mm. 
}
A spring mechanism with a diameter of \SI{45}{mm} is installed at the bottom of the inner cylinder. The inner cylinder was supported by the resistance of the springs at the base, and the stiffness of the inner cylinder when compressed from above could be controlled by varying the number of springs. 
\add{
\cref{fig:cylinder_c} shows five conditions for a cylinder spring mechanism. 
}
%図の左側には、シリンダーのスプリング機構の 5 つの定義済み剛性レベル (スプリング定数: \SI{0.5}{N/mm}) が示され、右側には機構の動的剛性調整プロセスの説明が示されています。
\del{
The right side of \cref{fig:spring} shows five conditions for a cylinder spring mechanism (spring constant: 0.5 N/mm), in order from left to right, as shown below.
}
%図4の右では次のように左から順番に円柱のばね機構（ばね定数：0.5 N/mm）の5条件を示す．
\del{
\begin{quote}
 \begin{itemize}
  \setlength{\parskip}{0mm} % 段落間
  \setlength{\leftskip}{20mm}
  \del{
  \item no springs
  \item 1 spring
  \item 2 springs
  \item 3 springs
  \item 4 springs
  }
 \end{itemize}
\end{quote}
}

The spring mechanism was designed such that the portion of the inner cylinder that compressed owing to the spring extended by \SI{15}{mm}. When the compression exceeded \SI{15}{mm}, the system behaved as a rigid body without any spring sensation. In the ``no spring'' condition, the inner cylinder functions as a rigid body from the outset because it is not equipped with springs. The lengths of both the outer and inner cylinders were carefully determined to ensure smooth movement of the inner cylinder within the outer cylinder and to prevent jamming even when the inner cylinder was compressed at an angle. Both the outer and inner cylinders were made of acrylic, a material chosen for its high sliding properties, ensuring that mobility is maintained even when the cylinders come into contact.
%使用するばね機構は，ばねによって伸縮する部分を\SI{15}{mm}になるように作成している．これは，ゴム関節が挿入された棒の曲がる変化量を基に設定している．\SI{15}{mm}以上に押し込む場合は，ばねの感覚が得られず剛体として機能する．「ばねなし」条件は，ばねを備えていないため，内筒を押し込み始める段階から剛体として機能する．また，外筒と内筒の長さは，外筒の内部で内筒がスムーズに移動できるように考慮し，内筒が斜めに押し込まれても詰まらないで降下できるよう検討して決定した．外筒および内筒は，各円筒同士が接触しても移動性が損なわれないように，摺動性の高いアクリル素材で作成した．

The hand postures used by humans when gripping objects can generally be categorized into two types: ``Precision Grip,'' where the object is grasped between the thumb and the flexed fingers, and ``Power Grip,'' where the object is held between the fingers and the palm while the thumb applies a counterforce~\cite{Napierhand}. Based on this understanding, this experiment adopted two conditions for gripping the stylus: Precision Grip and Power Grip.
%ここまでは，使用する道具の条件について述べたが，本実験では棒の持ち方についても条件を設定する．人間が物体を把持する際の手の姿勢は，物体を指の屈曲部と対向する親指で挟む込む形である「Precision Grip（精密把握）」と，物体を指と掌で挟み込んで親指で反力を加える形である「Power Grip（握力把握）」の2種類に大別できるとされている \cite{Napierhand}．この知見に基づき，本実験では棒の持ち方としてPrecision GripとPower Gripの2条件を採用した．

\subsection{Experimental Environment}\label{3_2}

\begin{figure}[t]
    \centering
    \includegraphics[width=80mm]{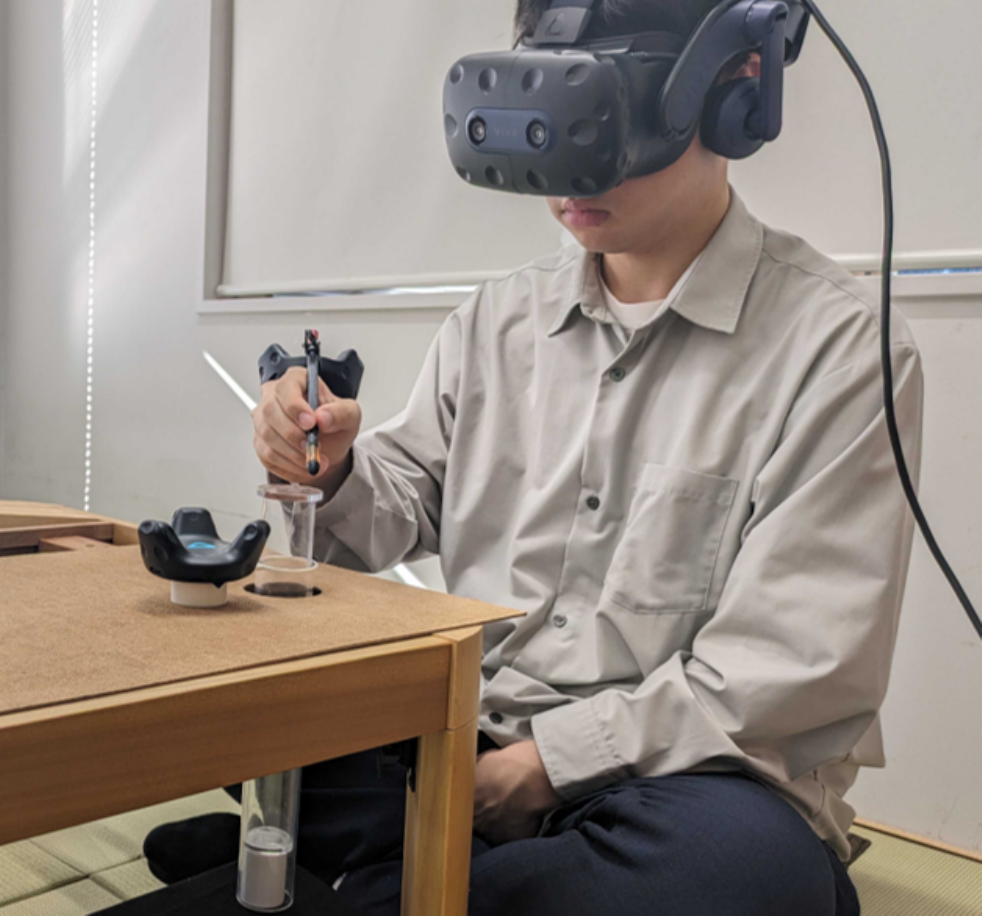} 
    %\vspace{-1cm}
    \caption{Experimental environment. Participants used the stylus to interact with a cylinder extending through a hole in the table, representing an object with variable stiffness.}
    \label{fig:vrjikken}
\end{figure}

\begin{figure*}[t]
    \centering
    \includegraphics[width=160mm]{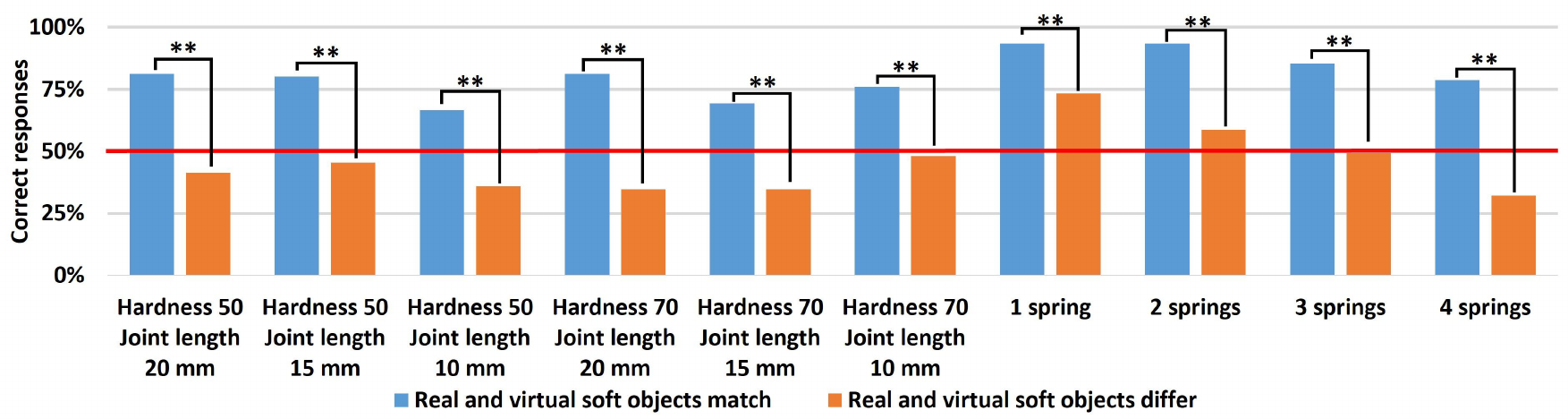}
    %\vspace*{-0.5cm}
    \caption{Probabilities of perception for each stylus/cylinder/visual condition in Precision Grip (**: $p<.01$).}
    \label{fig:pre1}
\end{figure*}

\begin{figure*}[t]
    \centering
    \includegraphics[width=160mm]{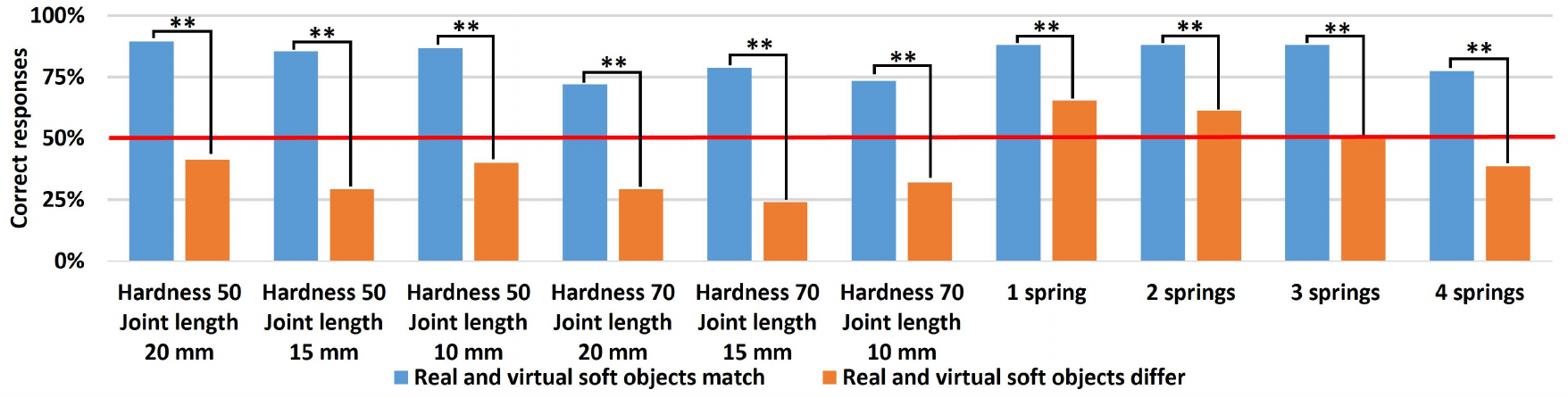}
    %\vspace*{-0.5cm}
    \caption{Probabilities of perception for each stylus/cylinder/visual condition in Power Grip (**: $p<.01$).}
    \label{fig:pow1}
\end{figure*}

\cref{fig:vrjikken} shows the experimental setup. VIVE trackers were attached to the bars and cylinders to reflect their positions and angles in virtual space. The table used in the experiment featured a hole with a diameter of \SI{50}{mm}, which allowed the cylinder to pass through. This design enabled participants to interact with the cylinder as an ``object with variable stiffness.'' The cylinder was extended 50 mm above the surface of the table. To eliminate auditory cues, the participants wore noise-canceling headphones (SONY WH-1000XM3) and were exposed to white noise.
%図\ref{fig:jikken}に実験環境を示す．棒と円柱にVIVEトラッカーを取り付けることで，現実の棒・円柱の位置と角度をバーチャル空間に反映させた．実験に使用したテーブルには，直径50 mmの穴を開けており，円柱を通すことができる．これにより，実験参加者が「硬さの変化する対象物」として円柱に触れることが可能になる．円柱はテーブルから50 mm突き出している．また，聴覚による手がかりを遮断する目的で，実験参加者にはノイズキャンセリングヘッドホン（SONY WH-1000XM3）を装着させ，ホワイトノイズを流した．

\subsection{Participants}\label{3_3}

Fifteen participants (8 males and 7 females, average age 22.4$\pm$0.9) were joined this experiment.
% undergraduate and graduate students at our university.
%at Aoyama Gakuin University. 
The participants were informed in advance that they would receive an honorarium based on the time spent in the experiment, which was conducted by touching rods and cylinders of various hardnesses while wearing the HMD, and that they could stop the experiment at any time.
%実験参加者は，青山学院大学の大学生および大学院生の男女計15名（男性8名，女性7名，平均年齢22.4$\pm$0.9 歳）であった．実験参加者にはあらかじめ，実験に要した時間に応じて所定の謝礼が支給されること，HMDを装着した状態で様々な硬さの棒と円柱に触れて評価する実験であること，実験はいつでも中止できることを伝え，同意書に署名をしてもらった．

\subsection{Procedure}\label{3_4}

The experiment consisted of two phases: a training session to familiarize participants with the procedure, and a test session to collect data. 
During the training session, the participants wore HMDs and observed the visual stimuli of deforming bars and cylinders. 
Once the participants indicated that they were sufficiently familiar with the procedure, the test session began. 
In the test session, participants still wearing the HMD were given a stylus with a VIVE tracker attached and were instructed to touch the cylinder. When the stylus was soft (except in the ``no rubber joint'' condition), the cylinder appeared hard (in the ``no spring'' condition). Conversely, when the stylus was hard (in the ``no rubber joint'' condition), the cylinder appeared soft (except in the ``no spring'' condition). Visual stimuli were presented in the virtual space for both the soft stylus and soft cylinder conditions.  During the measurement phase, participants used a two-alternative forced choice (2AFC)~\cite{2AFC} method to indicate whether the sensation of softness they experienced was caused by the stylus or cylinder. Participants were required to choose between the two options in each trial. The experiment included 200 trials— five trials for each of the 10 conditions (six stylus conditions and four cylinder conditions) and two visual stimulus conditions. 
The condition in which a stylus without a rubber joint touched a cylinder without a spring was excluded. 
The order of the stimulus presentation was randomized to mitigate any potential bias from the trial order.  After completing all the trials, the procedure was repeated under different stylus-holding conditions. Each participant's experiment lasted for approximately two hours.
%本実験は，実験参加者が実験方法に慣れるためのトレーニングセッションと，実際にデータを記録するためのテストセッションから構成される．
%トレーニングセッションでは，HMDを装着し，棒が変形する場合と円柱が変形する場合の視覚刺激を確認する．操作に十分慣れたと報告を受けた後，テストセッションへ移行した．テストセッションでは，HMDを装着した実験参加者に対し，VIVEトラッカーが接続された棒を渡して円柱に触れさせた．この際，棒が軟らかい場合（「ゴム関節なし」条件以外）は円柱が硬く（「ばねなし」条件），棒が硬い場合（「ゴム関節なし」条件）は円柱が軟らかく（「ばねなし」条件以外）なるようにする．現実で棒が軟らかい場合と円柱が軟らかい場合の両方で，バーチャル空間で棒が変形する場合と円柱が変形する場合の視覚刺激を提示する．測定では2AFCにより，実験参加者が感じた軟らかさの感覚が，棒によるものだったのか，円柱によるものだったのかを回答させた．また，必ず回答はどちらか1つで答えるように指示した．ここまでを1試行とし，本実験では，持ち方2条件で，硬い棒で硬い円柱に触れる条件を除外した6条件の棒と4条件の円柱である10条件，2条件の視覚刺激に対して各5回の計200試行を実施した．この際，実験順序によるバイアスを考慮して，各試行の提示順はランダムに設定した．すべての試行を終えた後，棒の持ち方の条件を変更して，上記の手順を再度繰り返した．一人当たりの実験時間は約2時間であった．
\subsection{Result}\label{3_5}

\subsubsection{Comparison of Perceived Stiffness Across Stylus and Cylinder Conditions}

\begin{figure}[t]
    \centering
    \includegraphics[width=80mm]{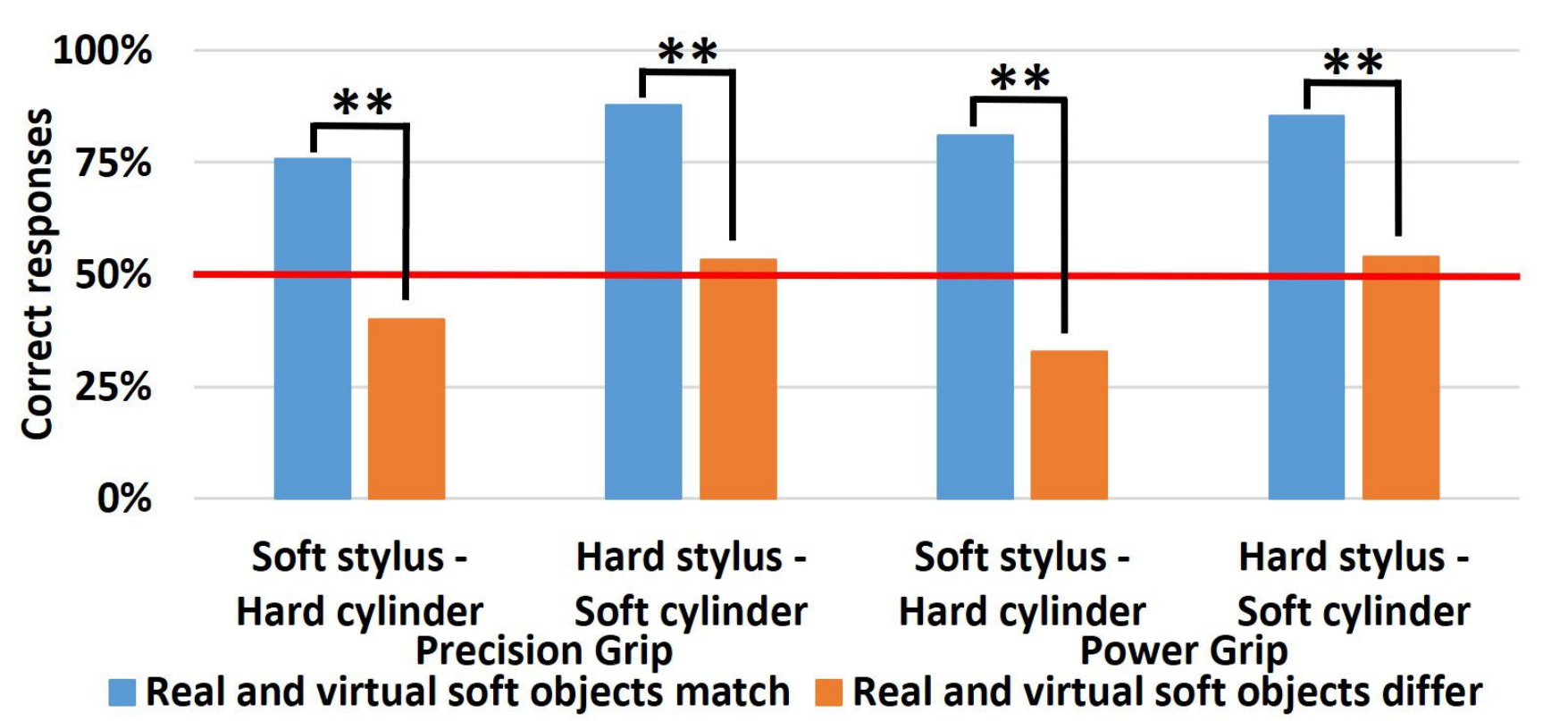} 
    %\vspace{-0.75cm}
    \caption{\add{Comparison of the correct answer rate when touching a hard cylinder with a soft stylus and when touching a soft cylinder with a hard stylus (**: $p<.01$).}}
    \label{fig:result2}
\end{figure}

The percentage of correct responses was calculated for each stylus, cylinder, and visual condition. The correct response rate in this experiment was defined as the participants’ ability to correctly perceive that the sensation of softness was caused by the stylus, such as when they touched a hard cylinder with a soft stylus. In other words, a higher percentage of correct responses (approaching 100\%) indicates that participants accurately judged whether the perceived softness was due to the stylus or cylinder under the given condition. 
\del{
Conversely, a percentage of correct responses closer to 0\% suggested that participants incorrectly perceived a hard object as soft under the same condition.  
}
\add{
To investigate the incidence of the illusion, we used bar graphs to directly compare correct response rates across conditions.
}
%錯覚の発生率を調査するために、棒グラフを使用して、条件間での正答率を直接比較しました。
To compare the correct response proportions across the stylus, cylinder, and visual conditions, a McNemar's test was conducted to determine the significant differences between the visual conditions. 
Additionally, $\varphi$(W), an indicator of the effect size in the McNemar's test, was examined. The $\varphi$ coefficient is considered small at 0.20, medium at 0.50, and large at 0.80~\cite{effect}.
\cref{fig:pre1} presents the percentage of correct responses for each stylus, cylinder, and visual condition in the Precision Grip condition, while \cref{fig:pow1} shows the corresponding data for the Power Grip condition.
%それぞれの棒および円柱，視覚条件ごとに，回答された結果の正答率を算出した．本実験での正答率とは，例えば，軟らかい棒で硬い円柱に触れた際に，実験参加者の感じた軟らかさが，正しく棒によるものだと感じたか否かを，正解・不正解と定義して求めている．つまり，正答率が100\%に近いほど，その条件において，実験参加者の感じる軟らかさが，棒と円柱のどちらに起因するものなのか正しく判断ができていると解釈できる．また，正答率が0\%に近いほど，その条件において，実験参加者の感じる軟らかさが，現実で硬い物体を軟らかく感じていると解釈できる．各棒および円柱，視覚条件における正答率について，対応のある比率の比較をするため，McNemar検定を実施し，各視覚条件間の有意差を調査した．このとき，McNemar検定における効果量の指標である$\varphi$（W）も確認した．$\varphi$係数は，0.20で小，0.50で中，0.80で大と判断する\cite{effect}．
%図\ref{fig:chapter4_1}に，Precision Gripでの各棒・円柱・視覚条件における正答率のグラフを示す．図\ref{fig:chapter4_2}に，Power Gripでの各棒・円柱・視覚条件における正答率のグラフを示す．

Real and virtual soft objects match when, for example, a user touches a cylinder with a soft stylus in reality, and the virtual environment presents a visual representation of the same soft stylus. 
In contrast, Real and virtual soft objects differ when the user touches a cylinder with a soft stylus in real life; however, the virtual environment presents a visual image of a soft cylinder instead.
%現実とバーチャルの軟らかい物体が一致というのは，例えば，現実で軟らかい棒で円柱に触れたときに，バーチャルでも軟らかい棒の視覚刺激を提示していることを示す．現実とバーチャルの軟らかい物体が異なるというのは，例えば，現実で軟らかい棒で円柱に触れたときに，バーチャルでは軟らかい円柱の視覚刺激を提示していることを示す．

Results showed that in both the Precision Grip and Power Grip, correct responses were higher when the real and virtual soft objects matched in all bar/cylinder conditions than when the soft objects differed. There was a significant difference in the percentage of correct responses between the visual conditions for all bar/cylinder conditions ($p\verb|<|0.01$).
%結果として，Presicoin GripとPower Gripのどちらも全ての棒・円柱条件で，現実とバーチャルの軟らかい物体が一致しているときは，軟らかい物体が異なるときよりも正答率が高くなった．視覚条件間では全ての棒・円柱条件で正答率に有意に差があった（p\verb|<|0.01）．

In the Precision Grip task, the percentage of correct responses exceeded 50\% for the ``1 spring'' and ``2 springs'' cylinder conditions when real and virtual soft objects differed. For other conditions, the percentage was below 50\%.
%Precision Gripでは，「ばね1つ」「ばね2つ」の円柱の条件で，現実とバーチャルの軟らかい物体が異なるときの正答率は50\%を超え，それ以外の条件は，50\%を下回っていた．
Among the visual conditions, the effect sizes were as follows:
$\varphi$ = 0.411 for ``Hardness 50 \& Joint length \SI{20}{mm},'' 
$\varphi$ = 0.358 for ``Hardness 50 \& Joint length \SI{15}{mm},'' $\varphi$ = 0.307 for ``Hardness 50 \& Joint length \SI{10}{mm},'' $\varphi$ = 0.473 for ``Hardness 70 \& Joint length \SI{20}{mm},''  $\varphi$ = 0.347 for ``Hardness 70 \& Joint length \SI{15}{mm},'' $\varphi$ = 0.288 for ``Hardness 70 \& Joint length \SI{10}{mm},''  
$\varphi$ = 0.268 for ``1 spring,'' 
$\varphi$ = 0.406 for ``2 springs,'' $\varphi$ = 0.384 for `3 springs,'' $\varphi$ = 0.469 for ``4 springs.'' 
These effect sizes were close to moderate. 
%視覚条件間で，「硬度50 \& 関節長\SI{20}{mm}」は$\varphi$=0.411，「硬度70 \& 関節長\SI{20}{mm}」は$\varphi$=0.473，「ばね2つ」は$\varphi$=0.406，「ばね4つ」は$\varphi$=0.469，「硬度50 \& 関節長\SI{15}{mm}」は$\varphi$=0.358，「硬度50 \& 関節長\SI{10}{mm}」は$\varphi$=0.307，「硬度70 \& 関節長\SI{15}{mm}」は$\varphi$=0.347，「硬度70 \& 関節長\SI{10}{mm}」は$\varphi$=0.288，「ばね1つ」は$\varphi$=0.268，「ばね3つ」は$\varphi$=0.384と効果量は中程度に近い値を示した．

In the Power Grip task, the correct response rate for the ``1 spring,'' ``2 springs,'' and ``3 springs'' cylinder conditions exceeded 50\% when the real and virtual soft objects were different, while the other conditions resulted in rates below 50\%. 
%Power Gripでは，「ばね1つ」「ばね2つ」「ばね3つ」の円柱の条件で，現実とバーチャルの軟らかい物体が異なるときの正答率は50\%を超え，それ以外の条件は，50\%を下回っていた．
Among the visual conditions, the effect sizes were as follows: $\varphi$ = 0.504 for ``Hardness 50 \& Joint length \SI{20}{mm},'' $\varphi$ = 0.566 for ``Hardness 50 \& Joint length \SI{15}{mm},'' $\varphi$ = 0.484 for ``Hardness 50 \& Joint length \SI{10}{mm},''  $\varphi$ = 0.427 for ``Hardness 70 \& Joint length \SI{20}{mm},''  $\varphi$ = 0.547 for ``Hardness 70 \& Joint length \SI{15}{mm},'' $\varphi$ = 0.414 for ``Hardness 70 \& Joint length \SI{10}{mm},'' $\varphi$ = 0.268 for ``1 spring,'' $\varphi$ = 0.307 for ``2 springs,'' $\varphi$ = 0.405 for ``3 springs,'' $\varphi$ = 0.392 for ``4 springs.'' These effect sizes were close to moderate. 
%視覚条件間で，「硬度50 \& 関節長\SI{20}{mm}」は$\varphi$=0.504，「硬度50 \& 関節長\SI{15}{mm}」は$\varphi$=0.566，「硬度50 \& 関節長\SI{10}{mm}」は$\varphi$=0.484，「硬度70 \& 関節長\SI{20}{mm}」は$\varphi$=0.427，「硬度70 \& 関節長\SI{15}{mm}」は$\varphi$=0.547，「硬度70 \& 関節長\SI{10}{mm}」は$\varphi$=0.414，「ばね3つ」は$\varphi$=0.405，「ばね1つ」は$\varphi$=0.268，「ばね2つ」は$\varphi$=0.307，「ばね4つ」は$\varphi$=0.392と効果量は中程度に近い値を示した．

These results indicate that the presentation of visual stimuli significantly affected the perception of stiffness. In all conditions, the effect sizes were medium to large, underscoring the notable influence of the visual stimuli. 
When a hard cylinder with fewer springs was touched with a hard stylus, a large percentage of the participants incorrectly perceived the cylinder as soft, even when the real and virtual objects differed in softness. 
Conversely, when a soft cylinder with more springs was touched with a hard stylus, participants perceived the stylus as soft more frequently than when they were given a visual stimulus for a soft stylus in virtual reality. 
In all conditions involving realistic soft styluses, the participants perceived the cylinders as soft when presented with a virtual soft stylus visual stimulus.
%これらの結果から，視覚刺激の提示が硬軟の知覚に影響を与えることが分かった．全ての条件で，効果量は中から大の値を示しているため，視覚刺激による効果は大きいと言える．ばねの数が少ない硬い円柱を硬い棒で触れる際には，現実とバーチャルの軟らかい物体が異なるときでも円柱が軟らかいことを正しく知覚できている割合が多いことが分かった．ばねの数が多い軟らかい円柱を硬い棒で触れる際に，バーチャルで軟らかい棒の視覚刺激を与えると，棒が軟らかいと感じる割合が多いことが分かった．また，全ての現実で軟らかい棒条件では，バーチャルで軟らかい円柱の視覚刺激を与えると，円柱を軟らかいと感じている割合が多いことが分かった．

\subsubsection{Effects of coincidence of visual conditions and haptic presentation on stiffness perception}

To compare the sensations of touching a hard cylinder with a soft stylus and a soft cylinder with a hard stylus, we calculated the percentage of correct responses for the stylus and cylinder conditions, respectively. 
McNemar's test was conducted to compare the proportions of correct responses between the two visual presentation conditions and determine significant differences between them. The left side of \cref{fig:result2} shows the percentage of correct responses in the Precision Grip condition for the stylus and cylinder conditions, whereas the right side shows the percentage of correct responses in the Power Grip condition for both the stylus and cylinder conditions.
%軟らかい棒で硬い円柱に触れたときと，硬い棒で軟らかい円柱に触れたときの感覚を比較するため，棒条件全体と円柱条件全体でそれぞれ正答率を算出した． また，視覚を提示した2条件間における正答率について，対応のある比率の比較をするため，McNemar検定を実施し，条件間の有意差を調査した．図\ref{fig:}の左にPrecision Gripにおける棒条件全体と円柱条件全体の正答率の結果を，右にPower Gripにおける棒条件全体と円柱条件全体の正答率の結果を示す．

In the visual condition where ``Real and virtual soft objects differ,'' the correct response rate when touching a hard cylinder with a soft stylus was 40\% in the Precision Grip and 33\% in the Power Grip. The correct response rate when touching a soft cylinder with a hard stylus was 53\% for Precision Grip and 54\% for Power Grip. A higher proportion of the participants mistakenly perceived the virtual soft cylinder as a real soft object. In both the Precision Grip and Power Grip conditions, there was a significant difference between the ``Real and virtual soft objects match'' and ``Real and virtual soft objects differ'' conditions when touching a hard cylinder with a soft stylus and when touching a soft cylinder with a hard stylus ($p\verb|<|0.01$). 
\del{
Across all visual conditions, the correct response rate was higher when a hard stylus was used to touch a soft cylinder than when a soft stylus was used to touch a hard cylinder, indicating that the actual soft object was more accurately perceived.
}
%「現実とバーチャルの軟らかい物体が異なる」の視覚条件において，軟らかい棒で硬い円柱に触れるときの正答率は，Precision Gripでは40\%，Power Gripでは33\%であった．硬い棒で軟らかい円柱に触れるときの正答率は，Precision Gripでは53\%，Power Gripでは54\%であった．バーチャルで提示している軟らかい円柱を実際の軟らかい物体だと錯覚している割合が多くなっている．Precision GripとPower Gripの両条件において，軟らかい棒で硬い円柱に触れるときと硬い棒で軟らかい円柱に触れるときの両方で，「現実とバーチャルの軟らかい物体が一致」と「現実とバーチャルの軟らかい物体が異なる」の間で有意に差があった（p\verb|<|0.01）．全ての視覚条件で，軟らかい棒で硬い円柱を触れるときよりも，硬い棒で軟らかい円柱に触れるときの方が，正答率が高く，実際の軟らかい物体を正しく知覚できていることが分かった．

These results demonstrate that when the visual representations of real and virtual soft objects are aligned, real soft objects are accurately perceived. In addition, when the visual stimuli of real and virtual soft objects differ, the virtual soft object can still be perceived as a real soft object. 
Furthermore, the illusion of perceiving a real hard object as soft is more likely to occur when a soft stylus touches a hard cylinder than when a hard stylus touches a soft cylinder.
%これらの結果から，現実とバーチャルの軟らかい物体が一致している視覚刺激を提示することで，現実の軟らかい物体を正しく知覚できることが分かった．また，現実とバーチャルの軟らかい物体が異なる視覚刺激を提示することで，バーチャルの軟らかい物体を実際の軟らかい物体であると知覚していることが分かった．加えて，硬い棒で軟らかい円柱に触れるときよりも，軟らかい棒で硬い円柱に触れるときの方が，現実で硬い物体を軟らかい物体に感じる錯覚が起きやすいことが分かった．

\subsection{Discussion}\label{3_6}

\begin{figure}[t]
    \centering
    \includegraphics[width=80mm]{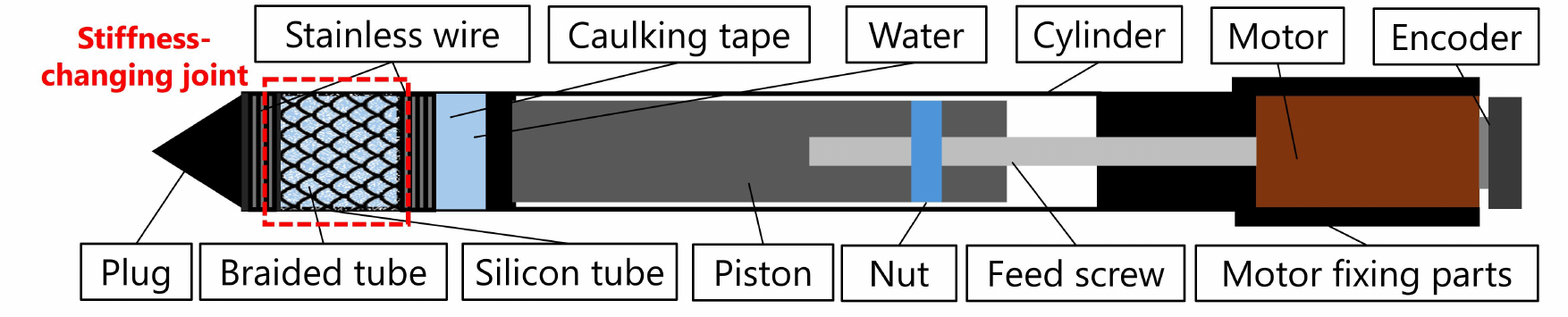}
    %\vspace*{-0.5cm}
    \caption{\add{Structure of the interface.}}
    \label{fig:structure}
\end{figure}
\begin{figure}[t]
    \centering
    \includegraphics[width=80mm]{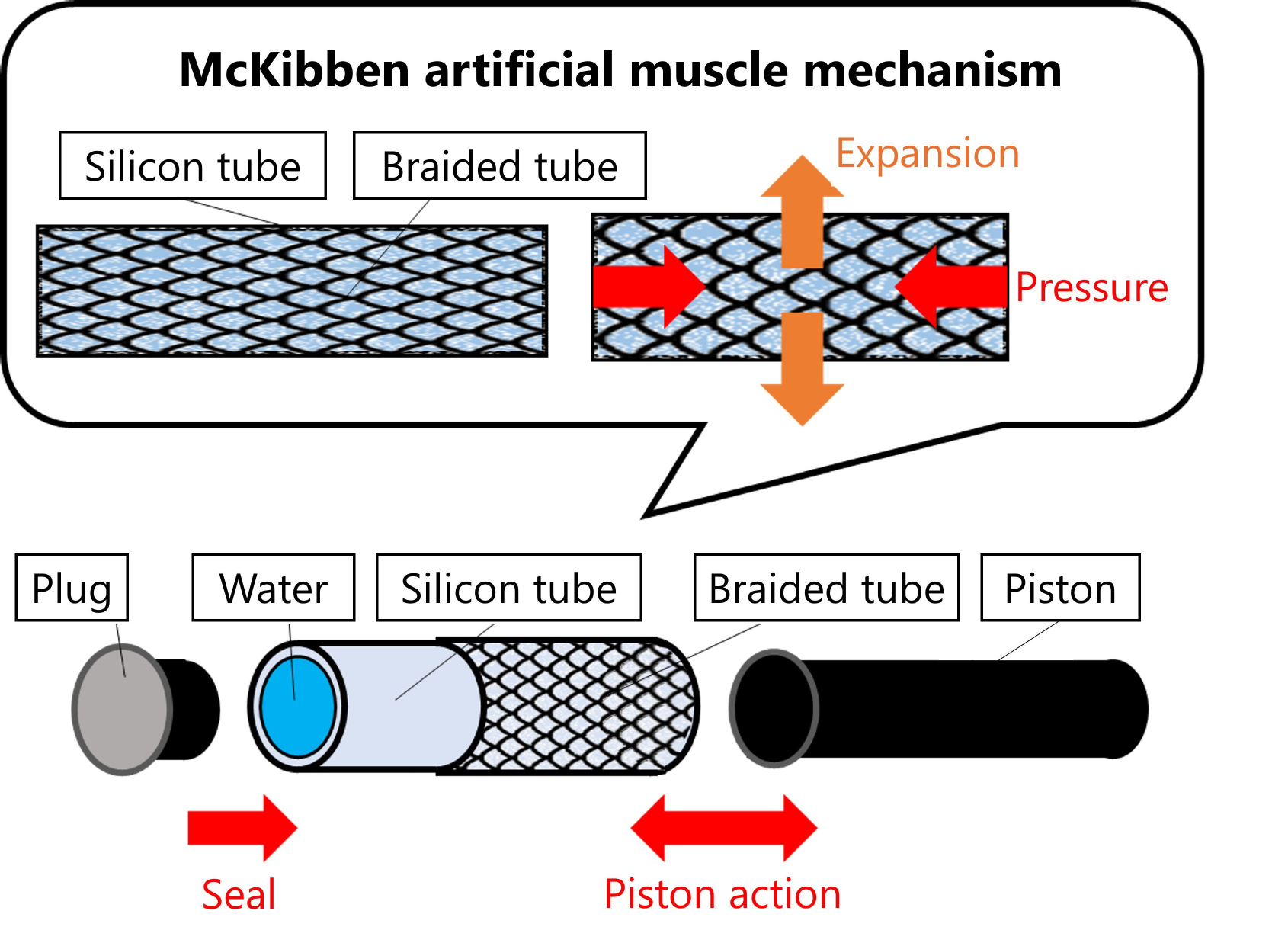}
    %\vspace*{-0.5cm}
    \caption{Mechanism of the stiffness change.
    %\add{The tube expands and becomes hard when pressure is applied to the McKibben-type artificial muscle.}
    }
    %マッキベン型人工筋肉に圧力をかけると、チューブが膨張して硬くなります。
    \label{fig:mechanism}
\end{figure}
\begin{figure}[t]
    \centering
    \includegraphics[width=80mm]{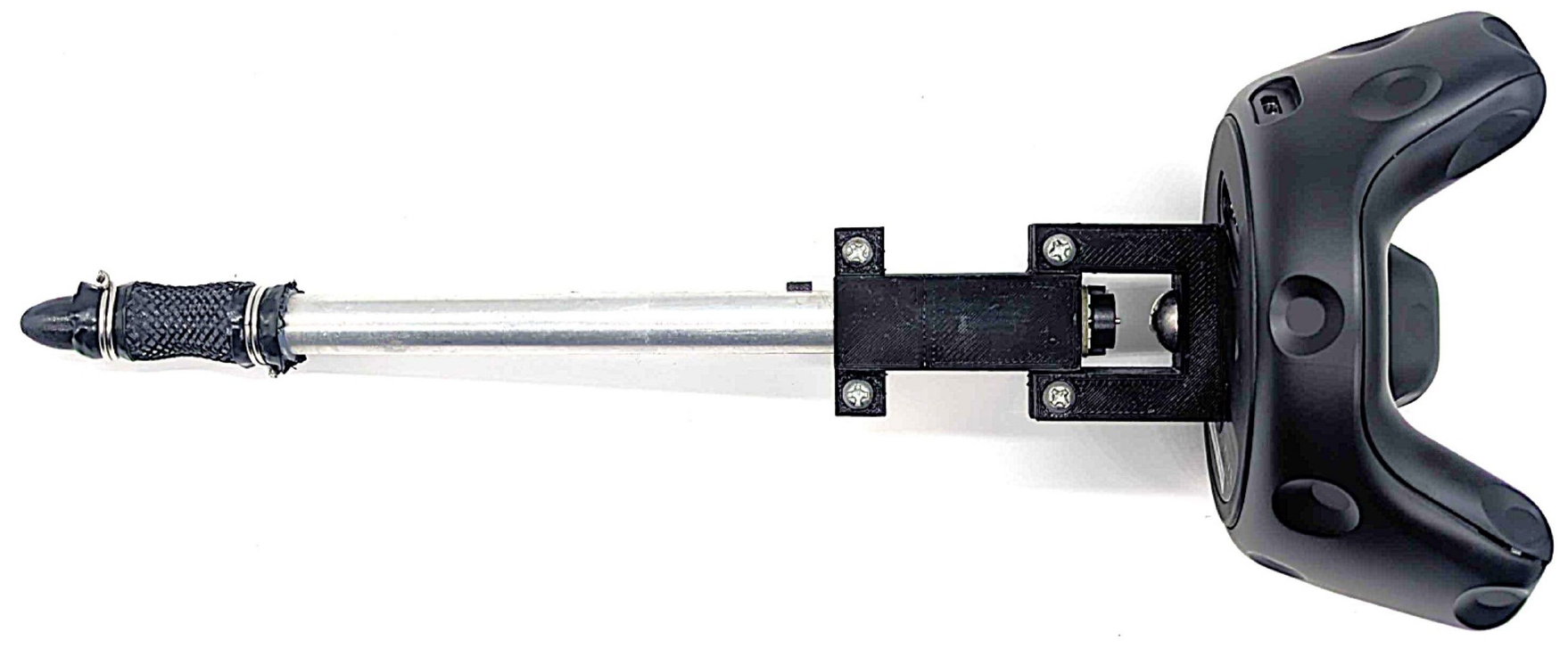}
    %\vspace*{-0.5cm}
    \caption{Appearance of the interface (joint length \SI{15}{mm}) with a VIVE Tracker.}
    \label{fig:Appearance}
\end{figure}

There was a significant difference in the percentage of correct responses between the visual conditions where ``real and virtual soft objects are the same'' and ``real and virtual soft objects are different.'' This may be because humans rely on both haptic and visual perception as key cues when perceiving object stiffness. The difference in the percentage of correct responses for the bar and cylinder conditions is likely due to the varying degrees of haptic and visual influences depending on the stiffness of the object. In the ``1 spring'' and ``2 springs'' conditions, the cylinders were too soft, leading to a stronger haptic influence than visual influence, which resulted in a higher percentage of correct responses.
%「現実とバーチャルの軟らかい物体が一致」と「現実とバーチャルの軟らかい物体が異なる」の視覚条件で正答率に有意に差があった．これは，人間が物体の硬軟を知覚する際には，触覚からの知覚だけでなく，視覚からの知覚も重要な手がかりにしているからだと考えられる．各棒・円柱の条件で正答率に違いがあるのは，物体の硬軟によって，触覚と視覚の影響力の大きさに違いがあるからだと考えられる．「ばね1つ」「ばね2つ」条件の円柱は，物体が軟らかすぎることで，視覚よりも触覚の影響を大きく受けたため正答率が高くなったのだと考えられる．

% \add{
% We set up the system such that the deformation of the virtual objects was visible in all stylus and cylinder conditions.
% }
%我々は，全ての棒・円柱条件で，バーチャルオブジェクトの変形が分かるように設定した．
\del{
To enhance the participants' perception of visual stimuli, the deformation of the virtual stylus and cylinder was exaggerated compared to that of a real soft object. 
}
%実験参加者が視覚刺激を認識しやすいようにするため，現実の軟らかい物体の変形よりもバーチャルの棒と円柱の変形を大きくしている．
Consequently, even under the ``Hardness 70,'' ``Joint length \SI{10}{mm},'' and ``4 springs'' conditions—where the stylus was stiffer and less deformable than in the bar and cylinder conditions—participants were still able to perceive the visual deformation and were equally influenced by the visual stimuli across all conditions.
%そのため，棒・円柱条件の中でも硬く変形しにくい「硬度70 \& 関節長\SI{10}{mm}」や「ばね4つ」の条件でも，実験参加者は物体が変形している視覚刺激を確実に認識しているため，他の条件と同等に視覚刺激による影響を受けていると考えられる．

\del{
The percentage of correct responses was lower when a soft stylus touched a hard cylinder than when a hard stylus touched a soft cylinder. This is likely because humans have limited experience in perceiving haptic sensations through soft styluses in everyday life, making it difficult for them to interpret such feedback accurately. 
Therefore, when visual stimuli were presented while the soft stylus was in contact with the hard cylinder, respondents appeared to rely more on visual perception to assess the stiffness of the object because haptic perception was less distinct.
}
%軟らかい棒で硬い円柱に触れるときが，硬い棒で軟らかい円柱に触れるときよりも正答率が低い．これは，人間は生活の中で軟らかい棒を介した触覚の経験が少ないため，知覚がしにくい．そのため，軟らかい棒で硬い円柱を触れたときに，視覚刺激が提示されていれば，知覚がしにくい触覚よりも視覚を頼りに物体の硬軟を知覚していたのだと考えられる．

These results indicate that presenting a virtual visual stimulus of a soft cylinder can create the illusion that a hard object felt soft when the user perceives haptic feedback via a soft stylus.
%以上の結果から，本実験を通して，軟らかい棒を介した触覚知覚をする際，バーチャルで軟らかい円柱の視覚刺激を提示することで，円柱が軟らかいという錯覚が起きる可能性を示せることが分かった．
% \add{
% Based on these findings, the interface will be designed in the following section to incorporate these parameters, ensuring the reproducibility and universality of the resulting phenomena.
% }
%これらの調査結果に基づいて、次の章ではこれらのパラメータを組み込むインターフェースを設計し、結果として得られる現象の再現性と普遍性を確保する。

\section{Implementation \& Measurement}
\label{4}

\begin{figure*}[t]
    \centering
    \includegraphics[width=\linewidth]{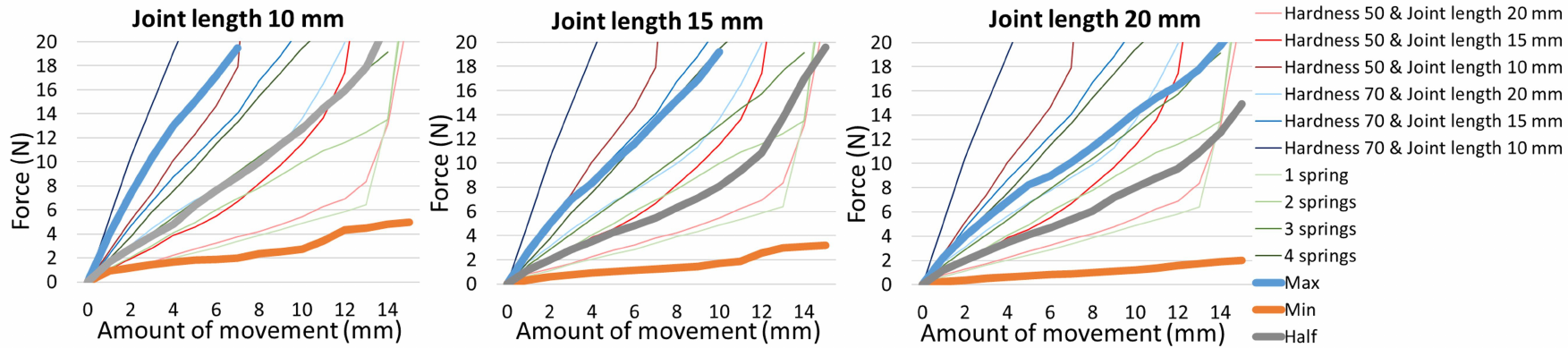}
    %\vspace*{-0.5cm}
    \caption{Average stiffness of \SI{10}{mm} \& \SI{15}{mm} \& \SI{20}{mm} joint length.}
    \label{fig:stylus_graph}
\end{figure*}

\add{
 The experiment described in \cref{3} demonstrates the possibility that touching a hard object with a soft stylus can create the illusion that the hard object felt soft.
 To replicate this phenomenon, we will construct an interface that can reproduce the measured stiffness properties of the soft stylus used in the experiment.
}
%3章の実験では、硬い物体に柔らかいスタイラスで触れると柔らかさの錯覚が生まれ、材質の硬さと触覚知覚の相互作用が強調されることが実証されています。これらの結果に基づいて、このセクションでは、3章で説明した柔らかいスタイラスの測定された剛性特性を再現する新しいインターフェースを紹介します。
% こうした現象を再現するインタフェースを実現するために、3章の実験で使用したスタイラスの測定された剛性特性を再現するインターフェースを構築します。

\del{
We found that the stiffness of the stylus used in the experiments (\cref{3}) could alter the perception of object stiffness. 
To utilize this finding for haptic rendering of stiffness of various objects, we aim to develop an interface capable of controlling the stylus stiffness.
}
%3章の実験で使用したスタイラスの剛性ではその錯覚を起こす可能性があることが分かった．そのため，我々は3章のスタイラスの剛性を提示できるインタフェースの作成を目指す．

\subsection{Implementation of Stylus-shaped Interface with Artificial Muscle}\label{4_1}

\cref{fig:structure} shows the structure of the interface. 
%図～ref{fig:structure}にデバイスの構造を示す。
\add{
We used McKibben artificial muscles to control stiffness of the interface.
The primary advantage of McKibben artificial muscles lies in their ability to control stiffness by modifying the length of the artificial muscle, thereby enabling macroscopic deformation. 
Given their use of silicone tubes, which allow for the control of stiffness, these muscles were hypothesized to effectively replicate the variable stiffness of the soft styluses used in \cref{3} with an inserted rubber element. 
}
%マッキベン人工筋肉の主な利点は、人工筋肉の長さを変更することで硬さを動的に調整し、マクロ的な変形を可能にすることです。硬さの調整を可能にするシリコンチューブを使用していることから、これらの筋肉はゴム要素が挿入された柔らかいスタイラスの可変硬さを効果的に再現すると想定されました。
\del{
The goal of this proposal is to alter the perceived stiffness of the objects touched by the interface by controlling the stiffness of the tube (joint) portion.
}
%本提案では，チューブ（関節）部分の硬軟を制御することで，デバイスで触れたものの硬さに対する感覚を変化させることを目標としている．
\cref{fig:mechanism} shows the mechanism of the stiffness change of the joint.
The appearance of the interface with a \SI{15}{mm} stiffness-controllable joint is shown in \cref{fig:Appearance}. This interface had a total length of \SI{250}{mm} and a weight of \SI{144}{g}, and a handle diameter of \SI{11}{mm}, including the VIVE tracker.
%\cref{}に剛性変化関節長が15mmのデバイスの外観を示す．関節長が15mmのデバイスは，全長が250mm，重量が144g，持ち手の径が 11 mmとなった(VIVE trackerを含む)．

Inside the interface, a feed screw connected to a DC motor with an encoder (DC motor, \SI{78}{rpm}) rotates to move the nut linearly, enabling the reciprocating motion of the attached piston. This motion pressurizes the liquid in the cylinder, pushing it toward the silicone tube located at the joint.  The joint utilizes a McKibben-type artificial muscle mechanism and consists of two layers: an inner silicone tube and outer polyester braided tube. When the piston pressurizes water at the joint, the silicone tube expands radially. However, the outer braided tube constrains this radial expansion, leading to axial contraction and an increase in the overall stiffness of the joint. \cref{fig:mechanism} illustrates the mechanism behind this increase in the stiffness.
%デバイス内部では，エンコーダ付きモータ（DCモータ，78 rpm）に接続された送りねじが回転することでナットが直線的に移動し，取り付けられたピストンの往復運動が可能となっている．これにより，シリンダに留まっていた液体が加圧されることでシリコンチューブ方向（関節部）に押し出される．関節部はマッキベン型人工筋肉の機構を備えており，内側がシリコンチューブ，外側がポリエステル製編組チューブの2層構造で構成されている．ピストンにより関節部で水を加圧すると，シリコンチューブは半径方向へ膨張しようする．しかし，外側の編組チューブが膨張を押さえつけるため，軸方向への収縮が起き，関節全体の剛性が増加する．図\ref{fig:figure3}に剛性が増加するメカニズムを示す．

Both ends of the tube were connected to the cylinder on the handle side, with a plug inserted at the tip to prevent liquid leakage. The plug and cylinder feature grooves, which allow the tube to securely cover both components. A stainless-steel wire was used to fasten the tube along the grooves. The cylinder was made of machined aluminum and a commercially available motor was used. The plug, piston, and motor mounting parts were 3D-printed.
%チューブの両末端は，液体の漏出を防ぐために，先端側ではプラグを挿入し，持ち手側ではシリンダに接続されている．プラグとシリンダには溝があることで，プラグとシリンダの上からチューブで覆い，ステンレスワイヤを用いて溝に沿って締結処理をしている．締結にはワイヤリング処理を採用している．シリンダはアルミパイプを加工して作成し，モータは一般的に販売されているものを使用している．また，プラグやピストン，モータ固定部品は3Dプリンタを利用して自作した．

The interface controls the direction and range of piston movement via a control board to control the stiffness and flexibility of the joint. The control board consists of a microcontroller (Arduino UNO) as the core of the interface, along with a motor driver (Toshiba TA7291P) to manage motor rotation. The direction of piston movement is controlled by reversing the rotation of the motor. The distance traveled by the piston was regulated by an encoder attached to the motor, which tracked the number of revolutions. The motor operates until the target number of revolutions is reached, thereby controlling the movement range. The stiffness of the interface was controlled using a PC control.
%本デバイスは，関節部の硬軟を変化させるため，制御基板を介してピストン運動の方向と移動距離を制御している．制御基板はデバイス制御の中心となるマイコン（Arduino UNO）と，モータの回転方向を変化させるためのモータドライバ（東芝製 TA7291P）から構成される．ピストン運動の方向は，モータの回転方向を切り替えることで制御する．一方で，ピストン運動の移動距離は，モータに取り付けられたエンコーダによって回転数を測定し，指定した回転数に到達するまでモータを駆動させる手法を取ることで，任意の距離分移動するよう制御する．私達はPCからの操作によりデバイスの剛性を制御している．

\subsection{Stiffness Measurement}\label{4_2}

\add{
To evaluate the capability of the interface for stiffness control, the stiffness values produced by the system were measured and compared to the stiffness range of the soft styluses used in \cref{3}. 
The comparison demonstrated that the interface outputs stiffness values within the target range, thereby achieving the desired control accuracy defined in this study. 
}
%動的剛性制御のためのインターフェースの能力を評価するために、システムによって生成された剛性値を測定し、第3章のソフトスタイラスの剛性範囲と比較しました。この比較により、インタフェースが目標範囲内の剛性値を出力し、研究で定義された目的の制御精度を達成していることが実証されました。
\del{
We quantitatively measured the change in the interface stiffness. 
}
To evaluate the effect of the different joint lengths, styluses with joint lengths of 10, 15, and \SI{20}{mm} were used.
We measured the average angles from the desk when the subjects naturally grasped the stylus using both a precision grip and a power grip, as described in \cref{3}. The average angles were $46^\circ$ for precision grip and $28^\circ$ for power grip. Next, we connected the stylus to a force gauge at a $46^\circ$ and measured the force required to bend the interface when vertically lowered. The tools used in \cref{3} were measured in the same manner and compared with these styluses.
%異なる関節長のデバイスを評価するために，関節長が10mm/15mm/20mmのデバイスを作成した．3章の実験で被験者が自然にスタイラスをprecision gripとpower gripで把持したときの机からの平均角度を測定し，precision gripでは46度，power gripでは28度であった．そこで，デバイスを46度に傾けた状態でフォースゲージに接続し，垂直に降下させたときに曲げるのに必要な力を測定した．また，3章で使用した道具も同様に測定し，デバイスと比較する．

Measurements were taken five times at each displacement as the force gauge was lowered from 0 to \SI{15}{mm}, and the average values were recorded. The piston displacement, from the maximum to the minimum, was divided into five segments. From these measurements, the piston displacement at which the interface could output half its maximum stiffness was determined. 
The maximum, half, and minimum stiffness of the interface at joint lengths of 10, 15, and \SI{20}{mm}, along with the measured stiffness of the tool described in \cref{3}, are shown in \cref{fig:stylus_graph}.
%デバイスと道具を1 mm間隔で，0 mmから\SI{15}{mm}にかけてフォースゲージを降下させた際の値を各5回取得し，その平均値を記録した．また，最大から最小までのピストンの移動量を5分割にして，そのときの値からデバイスが半分の剛性を出力できるピストンの移動量を求めた．\cref{fig:stylus_graph}に関節長が10mm/15mm/20mmでのデバイスの最大/半分/最小の剛性と3章の道具の剛性の測定結果を示す．

The results indicated that the minimum stiffness of each joint length was softer than that of any of the tools discussed in \cref{3}. For a joint length of \SI{10}{mm}, the maximum stiffness value fell between the configurations of ``Hardness 70 \& Joint length \SI{10}{mm}'' and ``Hardness 50 \& Joint length \SI{10}{mm}.'' For a joint length of \SI{15}{mm}, the maximum value was close to ``Hardness 70 \& Joint length \SI{15}{mm},'' while for a joint length of \SI{20}{mm}, the maximum was similar to ``Hardness 70 \& Joint length \SI{20}{mm}.'' These findings suggest that shorter joint lengths result in higher maximum stiffness, indicating a broader range of stiffness that the interface can provide.
%結果より，各関節長の最小値は3章のどの道具よりも軟らかいことが分かった．関節長が10㎜の最大値は「Hardness 70/Joiny length \SI{10}{mm}」と「Hardness 50/Joiny length \SI{10}{mm}」の間の硬さであった．関節長が15㎜の最大値は「Hardness 70/Joiny length \SI{15}{mm}」に近い硬さであった．関節長が20㎜の最大値は「Hardness 70/Joiny length \SI{20}{mm}」に近い硬さであった．また，デバイスの関節長は短いほど最大が硬く，提示できる剛性の範囲が広いことがわかる．

From the above findings, we conclude that this interface can control its stiffness to match that of the stylus used in \cref{3}. When combined with visual stimuli, the stylus in \cref{3} was able to create the illusion that a hard object felt soft. Therefore, we believe that this interface can similarly induce the same haptic illusion.
%以上より，本デバイスは剛性を変化させることができ，3章で使用したスタイラスに近い剛性を提示できることがわかった．3章のスタイラスは視覚刺激と合わせることで，硬い物体を軟らかい物体だという錯覚を起こす可能性を示せるので，同様に本デバイスでも十分にその錯覚を起こす可能性があると考えられる．

\section{Evaluation}
\label{5}

\del{
We conducted a user study to determine whether the interface could reproduce the stiffness of the virtual object material.}

\add{
This section shows an evaluation of whether the proposed interface can accurately simulate stiffness-based haptic feedback consistent with measurable deformations in the material properties of virtual objects.
}
%このセクションでは、\cref{3}の実験結果に基づいて開発され、\cref{4}で説明されているように実装されたインターフェースが、仮想オブジェクトの材料特性の測定可能な変形と一致する剛性ベースの触覚フィードバックを正確にシミュレートし、知覚的にリアルなユーザーエクスペリエンスを確保できるかどうかを評価します。
We prepared three different virtual object materials to assess the interface's ability to reproduce different virtual object materials by changing their stiffness.
\del{and to determine which of the three levels of stiffness of the interface is suitable for which material.}
%本章では，3段階の剛性のインタフェースを用いて3段階の剛性のマテリアルのバーチャルオブジェクトに触れることで，我々のインタフェースがバーチャルオブジェクトのマテリアルに適した触覚を提示できるのかを明らかにする．

\subsection{Participants}\label{5_3}

We recruited 12 participants (7 males and 5 females) aged between 21 and 24 years (Mdn = 22.4,
SD = 1.0). 
All participants had previously taken part in the experiment described in \cref{3}. 
The study aimed to investigate how the interface influences the user's haptic perception of stiffness. 
%N=12（男性7名、女性5名）の21歳から33歳（Mdn=27）のボランティア21歳から24歳のボランティア参加者（Mdn = 22.4、SD = 1)。 全員が第3章の実験にも参加した。我々は、この装置がユーザーの硬さの知覚にどのような影響を与えるかを調査した。

\subsection{Experimental Conditions \& Environment}

The evaluation used the interfaces with joint lengths of 10 and \SI{15}{mm}, capable of presenting a wide range of stiffness levels. The interfaces were tested under three conditions: maximum, half, and minimum stiffnesses.
%実験は提示できる剛性の範囲が広い関節長10ｍｍと15ｍｍのデバイスを使用した．デバイスの状態は最大/半分/最小の剛性の3条件で，
The virtual objects of contact represented different material stiffness levels: the stiffest was a tennis ball, the medium stiffness was a plastic ball (assumed to be made of polyvinyl chloride, PVC), and the softest was a sponge ball.
%バーチャルオブジェクトのマテリアルは最も硬いテニスボール，中間の硬さのプラスチックボール，最も軟らかいスポンジボールを用意した．プラスチックボールはポリ塩化ビニル製(PVC)を想定して作成した．
\add{
To ensure that the virtual interface stiffness corresponds appropriately to the observable macroscopic deformation of the simulated objects, we conducted a preliminary test similar to \cref{4_2} using a force gauge to measure and calibrate the material stiffness. 
Based on these tests, we selected a tennis ball to approximate the maximum stiffness of the interface, a plastic ball to approximate half the stiffness, and a sponge ball to approximate the minimum stiffness. 
}
%仮想インターフェースの剛性が、シミュレートされたオブジェクトの観察可能なマクロ的な変形に適切に対応していることを確認するために、フォースゲージ (セクション 4.2のような) を使用して材料の剛性を測定および較正する予備テストを実施しました。これらのテストに基づいて、インターフェースの最大剛性を近似するためにテニスボール、剛性の半分を近似するためにプラスチックボール、最小剛性を近似するためにスポンジボールを選択しました。この選択プロセスにより、各材料の物理的特性が異なる変形レベルに対応することが保証され、インターフェースのパフォーマンスを正確に評価できるようになります。
\add{
We configured the virtual environment to ensure virtual objects would deform when pressed with the stylus, using the same technique as described in \cref{3_1}. The deformation parameters were kept consistent across all materials.
}
%セクション 3.1 で説明した「vertexmotion」を使用して、スタイラスで押すと仮想オブジェクトが変形するように仮想環境を構成しました。変形パラメータはすべてのマテリアルにわたって一貫していました。
\del{
We selected these materials by investigating those with stiffnesses close to the three stiffness conditions of the interface. 
}
%これらのマテリアルはデバイスの剛性3条件に近い剛性を持つものを調査して選んだ．
\del{
We configured the virtual object to appear concave when the stylus was pressed into it in the VR environment. The concavity parameter is consistent across all materials.
}
%我々はVR空間内でバーチャルオブジェクトをスタイラスで押し込むとバーチャルオブジェクトが凹むように設定した．凹み具合のパラメータは各マテリアルで同じにしている．

The real object used in the experiment was a \SI{70}{mm} diameter sphere, 3D-printed, and attached to a VIVE tracker. 
Its size closely matched that of the actual material of the chosen virtual object.
%そのサイズは、選択した仮想オブジェクトの実際の素材のサイズとほぼ一致しました。 
%Its size closely matches that of the ball of the reference material.
%実験で使用した現実のオブジェクトは直径70mmの球体を3dプリンターで作成したものであり，VIVEトラッカーに接続されている．これは参考にした実際のマテリアルのボールに近いサイズにしている．

\subsection{Procedure}

\begin{figure}[t]
        \centering
        \includegraphics[width=80mm]{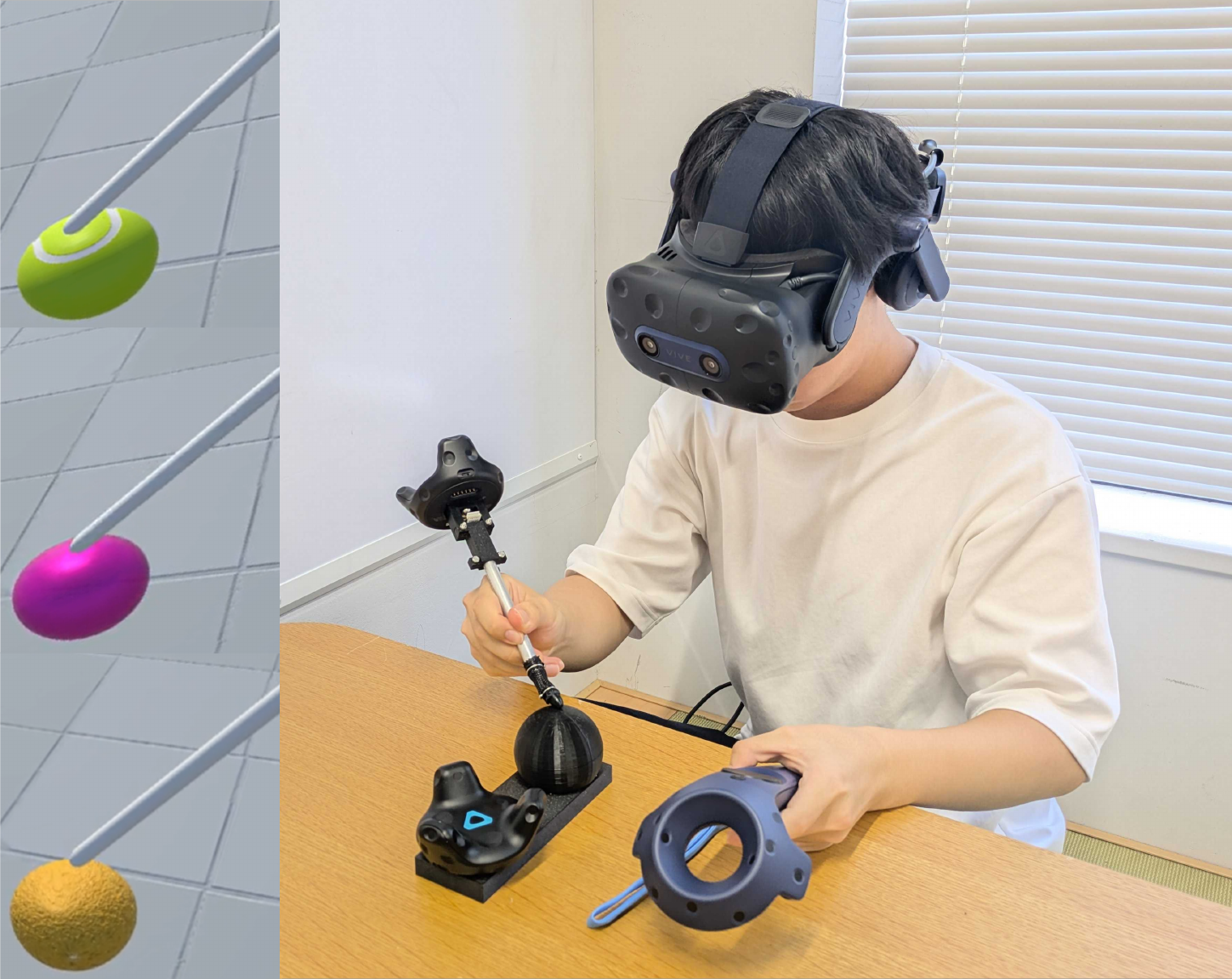}
    %\vspace*{-0.5cm}
        \caption{
        \add{Left: The virtual objects used in the evaluation, showing different material stiffness levels. Right: The participant touches the 3D-printed sphere with the interface.}
        \del{Left. The materials of the virtual objects, Right. The experimental setup}
        }
        \label{fig:userexperiment}
    \end{figure}

The experimental setup and materials of the virtual objects are shown in \cref{fig:userexperiment}. The participants wore an HMD (VIVE Pro 2) and grasped the stylus-shaped interface with their dominant hand using a precision grip. They touched the top of a virtual sphere with the interface in a pushing motion to receive haptic feedback. Participants were then asked to select the material that best matched the stiffness of the interface from the following three options: tennis, plastic, and sponge balls. Each participant completed 18 trials in total, with two conditions for the interface joint length, three conditions for the interface stiffness, and three trials for each condition. Finally, a free response questionnaire was administered.
%図??に実験環境を示す．実験参加者は，HMD(VIVE Pro 2)を装着し，デバイスを利き手でprecision gripで把持する．実験参加者は本デバイスで球体の上部を押し込むように触ることで触覚フィードバックを得る．渡されたデバイスの剛性にベストマッチするマテリアルをテニスボール，プラスチックボール，スポンジボールの3つ中から一つ回答する．デバイスの関節長2条件とデバイスの剛性3条件，試行回数3回により全部で18回回答してもらった．最後に自由回答のアンケート調査をした．
The order of presentation was randomized for each participant. To eliminate auditory cues, the participants wore noise-canceling headphones (SONY WH-1000XM3) and listened to white noise. The participants held a VIVE controller in their non-dominant hand and could change the material of the virtual object by pressing a button.
%各実験参加者の提示順序はランダムにした．また，聴覚による手がかりを遮断する目的で，実験参加者にはノイズキャンセリングヘッドホン（SONY WH-1000XM3）を装着させ，ホワイトノイズを流した．実験参加者はもう片方の手にVIVEコントローラーを持ち，ボタン操作によってバーチャルオブジェクトのマテリアルを自由に変更できる．

\subsection{Result}

\begin{figure}[t]
    \centering
    \includegraphics[width=80mm]{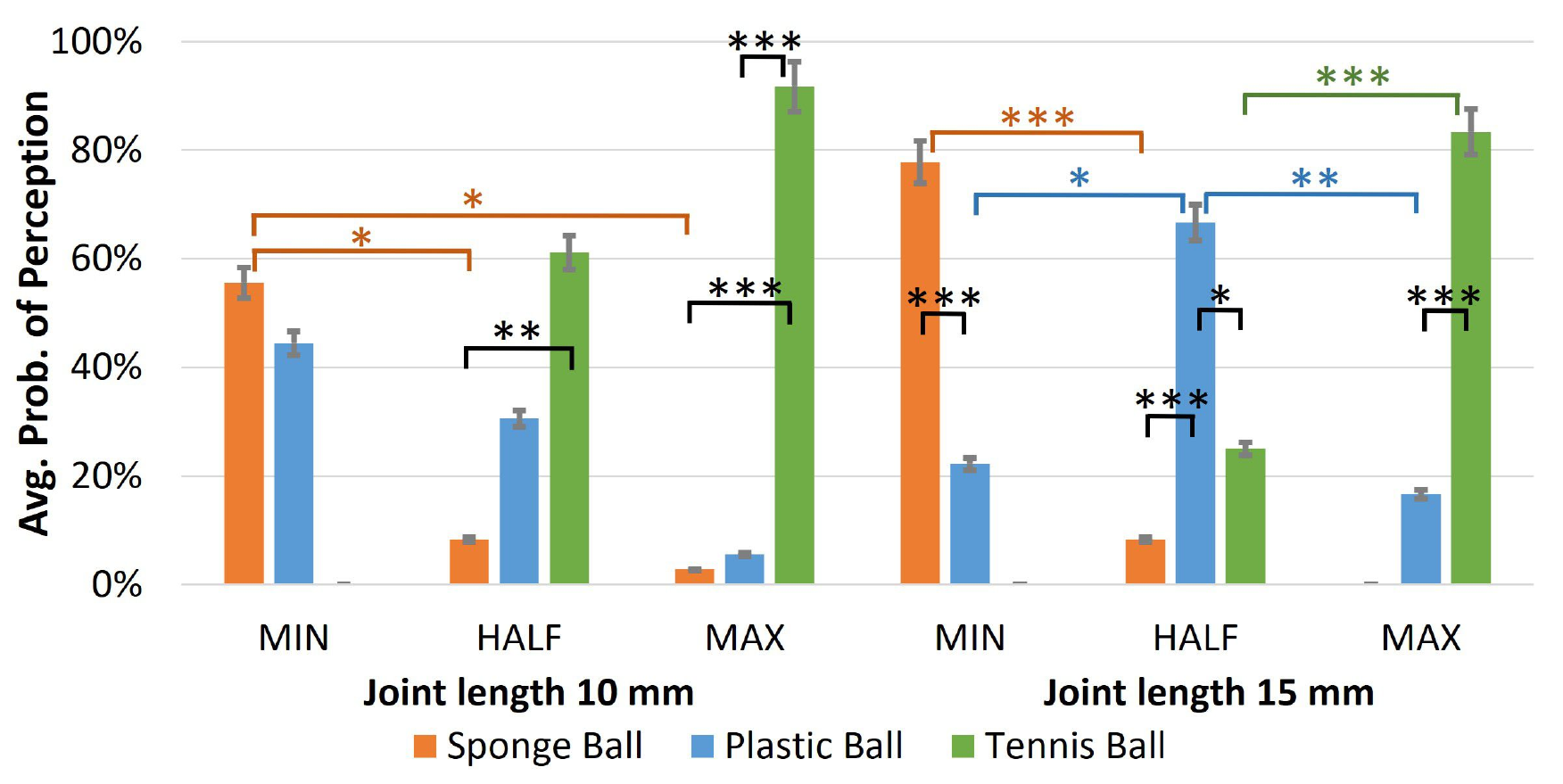}
    %\vspace*{-0.5cm}
    \caption{Probabilities of perceptions for the different materials. Brackets indicate statistically significant differences (*: $p<.05$; **: $p<.01$; ***: $p<.001$. Error bars show 95\% confidence intervals.}
    \label{fig:userresult}
\end{figure}

\begin{figure}[t]
    \centering
    \includegraphics[width=80mm]{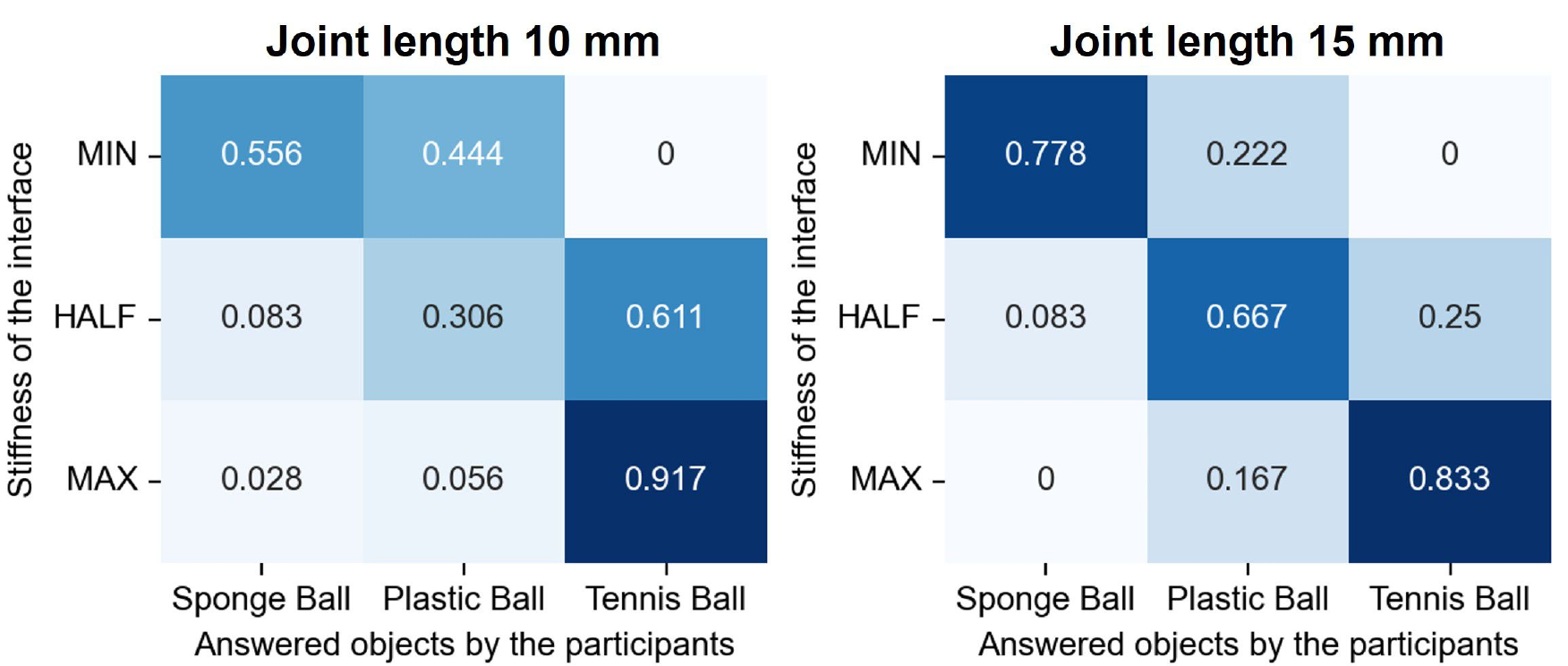}
    %\vspace*{-0.5cm}
    \caption{\add{
    %Confusion matrix derived from the results of the experiment. 
    Confusion matrix showing how participants identified the stiffness of the interface as corresponding to virtual objects in VR, for \SI{10}{mm} and \SI{15}{mm}.
    % In joint length \SI{15}{mm}, the percentage predicting correctly is much higher.
    Each row represents the ratio [\%] of responses for each stiffness level.}}
    %インタフェースの関節長が10mmと15mmの場合の実験結果から導出された混同行列。関節長15mmでは、正しく予測する割合が大幅に高くなります。各行は、VR でオブジェクトとして回答されたインターフェースの剛性の比率[%]を示しています。
    \label{fig:confusion}
\end{figure}

We calculated the perceived probability of each material that participants identified as the best match under the two interface joint length conditions and three interface stiffness conditions. \cref{fig:userresult} shows a graph of the results.
%デバイスの関節長2条件とデバイスの剛性3条件で，実験参加者がベストマッチだと回答した各マテリアルの知覚確率を算出した．図??に結果のグラフを示す．
Generalized linear mixed models (GLMM) with Tukey's HSD were used for multiple comparisons of the perceived probability of each material, accounting for the repeated responses of the experimental participants. We examined significant differences between conditions. 
 %各マテリアルの知覚確率について，実験参加者の複数回答を考慮して多重比較するために，一般化線形混合モデル(GLMM)をTukey's HSDで行った．我々は各条件間の有意差を調査した．

The analysis showed that, with a joint length of \SI{10}{mm}, the probability of perceiving the object as a ``tennis ball'' at the minimum interface stiffness was 0\%. For the \SI{15}{mm} joint length, the probability of perceiving the object as a ``tennis ball'' at the minimum stiffness and as a ``sponge ball'' at the maximum stiffness were both 0\%, with no significant differences due to zero variance.
%解析の結果，関節長10ｍｍでは，デバイスの剛性が最小のときにテニスボールと回答した知覚確率は0%であった．関節長15ｍｍでは，デバイスの剛性が最小のときにテニスボールと回答した確率とデバイスの剛性が最大のときにスポンジボールと回答した確率は0%であったため，分散が0であることからこれらと比較する際に有意に差がない．

For a joint length of \SI{10}{mm}, the probability of selecting ``tennis ball'' was highest when the interface stiffness was either at its maximum or at half its maximum. Conversely, when the interface stiffness was at its minimum, the probability of selecting a ``sponge ball'' was highest.
 %関節長10ｍｍでは，デバイスの剛性が最大のとき，テニスボールと回答している確率が最も高く，デバイスの剛性が半分のときにも，テニスボールと回答している確率が最も高かった．デバイスの剛性が最小のときは，スポンジボールと回答している確率が最も高かった．
There was a significant difference in the probability of selecting ``tennis ball'' compared to ``plastic ball'' or ``sponge ball'' when the interface's stiffness was at its maximum ($p\verb|<|0.001$). A significant difference was also observed between the probability of selecting ``tennis ball'' and ``sponge ball'' when the interface's stiffness was at half ($p\verb|<|0.01$). 
Additionally, the probability of selecting a ``sponge ball'' differed significantly when the interface had minimum stiffness compared with when it had maximum or half stiffness ($p\verb|<|0.05$).
 %デバイスの剛性が最大のときにテニスボールと回答した確率とプラスチックボールやスポンジボールと回答した確率には有意に差があった（p\verb|<|0.001）．デバイスの剛性が半分のときにテニスボールと回答した確率とスポンジボールと回答した確率には有意に差があった（p\verb|<|0.01）．デバイスの剛性が最小のときにスポンジボールと回答した確率と，デバイスの剛性が最大/半分のときにスポンジボールと回答した確率には有意に差があった（p\verb|<|0.05）．

For a joint length of \SI{15}{mm}, the probability of selecting ``tennis ball'' was highest when the interface stiffness was at its maximum. The probability of selecting ``plastic ball'' was highest when the interface stiffness was at its half. The probability of selecting ``sponge ball'' was highest when the interface stiffness was at its minimum.
%関節長15mmでは，デバイスの剛性が最大のとき，テニスボールと回答している確率が最も高く，デバイスの剛性が半分のときにも，プラスチックボールと回答している確率が最も高く，デバイスの剛性が最小のときは，スポンジボールと回答している確率が最も高かった．
There was a significant difference in the probability of selecting ``tennis ball'' compared to ``plastic ball'' when the interface's stiffness was at its maximum ($p\verb|<|0.001$).
%デバイスの剛性が最大のときにテニスボールと回答した確率とプラスチックボールと回答した確率には有意に差があった(p<0.001)．
Furthermore, the probability of answering the ``tennis ball'' was significantly different when the interface stiffness was at its maximum and half ($p\verb|<|0.001$).
%さらにテニスボールと回答した確率はデバイスの剛性が最大のときと半分のときに有意に差があった(p<0.001)．
When the stiffness of the interface was half, there was a significant difference between the probability of answering ``plastic ball'' and the probability of answering ``tennis ball'' ($p\verb|<|0.05$) or ``sponge ball'' ($p\verb|<|0.001$).
%interfaceの剛性が半分のときにプラスチックボールと回答した確率とテニスボール(p<0.05)，スポンジボール(p<0.01)と回答した確率には有意に差があった．
Furthermore, the probability of responding to a ``plastic ball'' was significantly different when the interface was half as stiff as when it was at maximum ($p\verb|<|0.01$) or minimum ($p\verb|<|0.05$) stiffness.
%さらにプラスチックボールと回答した確率はinterfaceの剛性が半分のときと最大(p<0.01)または最小(p<0.05)のときに有意に差があった．
Finally, there was a significant difference in the probability of selecting ``sponge ball'' compared to ``plastic ball'' when the interface's stiffness was at its minimum ($p\verb|<|0.001$). 
The probability of selecting a ``sponge ball'' also showed a significant difference when comparing the minimum stiffness of the interface to its half-stiffness setting ($p\verb|<|0.001$).
%デバイスの剛性が最小のときにスポンジボールと回答した確率とプラスチックボールと回答した確率には有意に差があった(p<0.001)．さらにスポンジボールと回答した確率はデバイスの剛性が最小のときと半分のときに有意に差があった(p<0.001)．

\add{
\cref{fig:confusion} shows the confusion matrix illustrating how different joint lengths and stiffness levels affected users’ ability to identify virtual object materials in VR. With a joint length of \SI{15}{mm}, the system successfully distinguished between the three levels of stiffness in the VR environment, as confirmed by the evaluation results. However, with a joint length of \SI{10}{mm}, the system was unable to differentiate between the sponge ball and the plastic ball when the stiffness was at its lowest level and was unable to successfully distinguish between the three stiffness levels in the VR environment.
%\cref{fig:confusion} は、さまざまなジョイント長と剛性レベルが VR で仮想オブジェクトの材質を識別するユーザーの能力にどのように影響するかを示す混同マトリックスを示しています。ジョイント長が \SI{15}{mm} の場合、システムは VR 環境の 3 つの剛性レベルを正常に区別し、評価結果で確認されました。ただし、ジョイント長が \SI{10}{mm} の場合、システムは剛性が最低レベルのときにスポンジボールとプラスチックボールを区別できず、VR 環境の 3 つの剛性レベルを正常に区別できませんでした。
%\cref{fig:confusion} shows the confusion matrix that illustrates the impact of different joint lengths and stiffness conditions on users’ ability to distinguish virtual object materials in VR. For an interface with a joint length of \SI{15}{mm}, the proposed haptic perceptual model accurately differentiated three distinct levels of stiffness in the VR environment, as validated by user performance metrics. However, for the \SI{10}{mm} joint length, the system failed to distinguish between the sponge ball and the plastic ball at minimum stiffness. 
}
%図は、実験結果から得られた混同行列を示しています。ジョイント長が \SI{15}{mm} のインターフェースの場合、提案された触覚知覚モデルは、ユーザー パフォーマンス メトリックによって検証されたように、VR 環境内の 3 つの異なるレベルの剛性を正確に区別しました。ただし、ジョイント長が \SI{10}{mm} の場合、システムは最小剛性のスポンジ ボールとプラスチック ボールを区別できませんでした。
 
\subsection{Discussion}

When the joint length is \SI{10}{mm}, the probability of responding ``tennis ball'' increases when the interface stiffness is set to half of the maximum, while the probability of responding ``sponge ball'' increases when the interface stiffness is at its minimum, rather than ``plastic ball,'' which represents the mid-range stiffness. These results suggest that interfaces with a joint length of \SI{10}{mm} do not provide sufficient haptic feedback to accurately convey the virtual object material, even when the stiffness of the interface is controlled. This may be because of the short joint length, which limits the concavity of the virtual object, making it difficult to perceive the intended haptic sensation.
%関節長が10mmのときに，デバイスの剛性が最大と半分のときにテニスボールと回答した確率が高くなっており，デバイスの剛性が最小のときに中間の硬さであるプラスチックボールではなく，スポンジボールと回答した確率が高くなっている．この結果から，関節長が10ｍｍのデバイスではデバイスの剛性を変えてもバーチャルオブジェクトのマテリアルの触覚フィードバックが十分にできていないと考えられる．この原因として，関節長が10ｍｍだと短いため，バーチャルオブジェクトを十分に凹ませることができず，触覚を感じにくくなったと考えられる．

In contrast, when the joint length was \SI{15}{mm}, respondents were more likely to perceive the object as a ``tennis ball'' at maximum interface stiffness, a ``plastic ball'' at medium stiffness, and a ``sponge ball'' at minimum stiffness. These results suggest that a joint length of \SI{15}{mm} allows sufficient indentation of the virtual object and that varying the stiffness of the interface provides haptic feedback appropriate to the perceived material of the object.
%対して，関節長が15ｍｍときは，デバイスの剛性が最大のときにテニスボールと回答し，デバイスの剛性が半分のときにプラスチックボールと回答し，デバイスの剛性が最小のときにスポンジボールと回答する確率が高くなっている．この結果から，関節長が15ｍｍのデバイスでは，バーチャルオブジェクトを十分に凹ませることができ，デバイスの剛性を変えることでバーチャルオブジェクトのマテリアルに適した触覚をフィードバックできると考えられる．

\section{Overall Discussion \& Future Work}

\subsection{
\add{Influence of Stylus Stiffness and Joint Length
%Discussion on Stylus in Haptic Feedback
}
\del{Discussion on Stylus Stiffness and Joint Length in Haptic Feedback}
}
%触覚フィードバックにおけるスタイラスの剛性とジョイント長に関する議論
Although the experiments in \cref{3} demonstrated that all stylus could produce the illusion of softness when combined with visual stimuli, it remains unclear whether stylus with stiffness outside the tested range or with longer joint lengths can achieve the same effect. Therefore, we propose that further experiments using stylus with greater or lesser stiffness than those tested in this study will offer valuable insights into haptic perception across a broader range of stylus stiffnesses.
%3章の実験では視覚刺激と合わせることで，全てのスタイラスで硬い円柱が軟らかいという錯覚を起こす可能性を示したが，今回の実験で使用した範囲外の剛性のスタイラスやスタイラスの関節長の長さによってはその錯覚を起こすことが可能かどうかは不明である．そのため，今回の実験で使用したスタイラスよりも硬いもしくは軟らかいスタイラスを使用して実験することで，幅広い範囲の剛性のスタイラスでの触覚知覚の知見を得られると考えられる．

In \cref{4}, shortening the length of the stiffness-changing joint increased the maximum stiffness of the interface but reduced its bendable displacement. Conversely, lengthening the joint decreases the maximum stiffness while increasing the bendable displacement. A larger displacement allows for deeper depression of the virtual objects. Therefore, controlling the joint length of the interface based on the reproduced virtual object is expected to provide more suitable haptic feedback.
%4章のデバイスの実装について，剛性変化関節の長さを短くするとデバイスの最大剛性は大きくなるが，デバイスを曲げることができる変位量は小さくなる．また，剛性変化関節の長さを長くするとデバイスの最大剛性は小さくなるが，デバイスを曲げることができる変位量は大きくなる．デバイスを曲げることができる変位量が大きいほどバーチャルオブジェクトを深く凹ませることができる．そのため，再現したいバーチャルオブジェクトによってデバイスの関節長を変えることでより適した触覚をフィードバックできると考えられる．

\del{
The method proposed in this research for presenting the stiffness of a virtual object's material requires both a interface and physical object to be touched by the interface. Therefore, depending on the virtual object being reproduced, it may be necessary to prepare a physical object with the corresponding shape and stiffness. However, if the interaction is limited to touching the object from above, we believe that haptic feedback can be achieved using a flat surface, such as a desk, in combination with virtual imagery, without the need for an object of the exact shape.
}
%本研究で提案するバーチャルオブジェクトの素材の剛性提示方法では，デバイスだけでなく，デバイスで触れるための対象物が必要となる．そのため，再現したいバーチャルオブジェクトによって，その形の硬い物体を用意する必要がある．しかし，オブジェクトを上から触れるだけに限定するならば，正確にその形の物体を用意しなくても，机などの平面の物体があれば，バーチャルの映像と合わせて机に触れることで，触覚フィードバックが可能である可能性があると私たちは考える．

\subsection{
\add{Limitations and Future Directions}
\del{Implications of Stylus Stiffness on Tool-Mediated Haptic Perception and Interface Design}}
%スタイラスの硬さがツールを介した触覚知覚とインターフェース設計に与える影響

%インタフェースのリミテーションについて
\add{
While the proposed interface demonstrates promising results in enhancing haptic perception, several limitations have been identified.
}

\add{
First, the proposed interface was primarily designed for use with precision and power grips, assuming a bending motion as the primary mode of interaction. 
Although these standard stylus grips allow users to apply pressure naturally and intentionally, the effects of alternative interaction methods, such as pushing or compressing the tip, remain unclear. 
Future studies should explore whether similar perceptual effects can be observed with other modes of interaction, expanding the versatility of the interface.
% Despite its effectiveness, the proposed interface has certain limitations observed during the user evaluations. 
% One notable issue was that participants occasionally attempted to bend the stylus tip while interacting with the virtual objects. 
% Standard stylus grips, such as precision and power grips, allow users to apply pressure to objects in a natural and intentional manner. 
% However, other interaction methods with the proposed interface other than pressing, may require further exploration. 
% The investigation of more complex interactions would be a promising direction for future research.
}
%使用中に観察された 1 つの制限は、ユーザーが操作中にスタイラスの先端を曲げようとすることがあることです。ただし、精密グリップやパワー グリップなどの標準的なスタイラス グリップを使用すると、ユーザーは自然で意図した方法でオブジェクトに圧力をかけることができます。代替のインタラクション方法は限られていますが、より複雑なインタラクションの探求は、今後の研究の方向性として検討します。

%インタフェースの重さについて
\add{
Second, although none of the participants reported discomfort or performance issues related to the device weight in the evaluation (\cref{5}), our work did not investigate the impact of different weight configurations.
This aspect represents a limitation, as the optimal weight for usability and performance remains undetermined. 
Future research should compare devices with varying weights to provide a more comprehensive understanding of how weight influences user experience.
% suggesting that its impact was minimal in this context. However, our study did not compare weight differences across various devices, which is a limitation. 
% Future studies should address this gap to provide a comprehensive understanding of the influence of weight on usability and performance.
}
%さらに、評価においてデバイスの重量に関連する不快感やパフォーマンスの問題を報告した参加者はいなかった (\cref{5})。これは、この状況ではデバイスの重量の影響は最小限であったことを示唆している。ただし、この研究ではさまざまなデバイスの重量の違いを比較していないため、限界がある。今後の研究では、このギャップを解消し、重量が使いやすさとパフォーマンスに与える影響を包括的に理解できるようにする必要がある。

%インタフェースの内部圧力について
\add{
Third, another area for improvement is the relationship between the internal pressure and interface stiffness. 
While this relationship is essential for understanding and optimizing haptic feedback, the structural limitations of the current interface prevent the direct measurement of the internal pressure. 
Overcoming these limitations and integrating such measurements into future iterations of the interface will significantly enhance the precision of the stiffness control.
}
%改善すべきもう 1 つの重要な領域は、内部圧力とインターフェースの剛性の関係です。この関係は触覚フィードバックを理解して最適化するために不可欠ですが、現在のインターフェースの構造上の制限により、内部圧力を直接測定することはできません。これらの制限を克服し、このような測定をインターフェースの将来の反復に統合することで、剛性制御の精度が大幅に向上します。

%モーターのトルクと速度について
\add{
Finally, the motor used in this study was selected because of its high torque at a rated voltage of \SI{12}{V} and rotational speed of \SI{78}{rpm}. However, a trade-off exists between the torque and rotational speed. Although a higher rotational speed is necessary for real-time stiffness controls, sufficient torque is also crucial. Balancing these factors is essential for developing robust and responsive systems.
}
%さらに、本研究で使用したモーターは、定格電圧 12 V、回転速度 78 rpm で高いトルクが得られることから選択されました。ただし、トルクと回転速度の間にはトレードオフが存在します。リアルタイムの剛性調整には高い回転速度が必要ですが、十分なトルクも重要です。これらの要素のバランスをとることは、堅牢で応答性の高いシステムを開発するために不可欠です。

\del{
We believe that the results of this study can be extended to other areas of tool-mediated haptic research. This study revealed that the stiffness of the gripping stylus influences the haptic perception of the object being touched through the stylus. These findings suggest that the stiffness properties of materials and structures in experimental apparatuses and interfaces must be carefully considered when investigating haptic perception and when designing haptic interfaces. For example, in tool-mediated haptic perception studies, the stiffness of tools used should be treated as an experimental parameter. Likewise, in the design of haptic feedback interfaces, it is essential to create a comprehensive haptic experience by factoring the stiffness of the interface.
}
%本研究で得られた結果は，他の道具を介した触覚研究にも波及すると考える．本研究を通して，把持する棒の硬軟が棒を介して触れる物体の硬軟知覚に影響があることが分かった．こうした知見は，触覚知覚の調査や触覚デバイスの設計において，実験器具やデバイスの素材や構造を由来とする硬軟を慎重に考慮する必要があることを示唆している．例えば，道具を介した触覚の調査においては，実験で使用する道具の硬軟を実験のパラメータとして考慮したり，触覚提示デバイスの設計においては，デバイスの硬軟を考慮した上で包括的に触覚体験をデザインしたりする必要性がある．

% \subsection{Prospects with Improved Interface Performance}
\del{
\subsection{
%\add{Implications for Tool-Mediated Haptic Research and Interface Design}
Improvement of Interface Performance}
}
% \add{
% The findings of this study extend beyond the immediate context and offer valuable insights into broader applications in tool-mediated haptic research and interface design. Our results underscore the importance of considering the stiffness properties of the materials and structures in both experimental setups and haptic interfaces. For instance, in studies involving tool-mediated haptic perception, the stiffness of the tools used should be treated as an experimental parameter, because it significantly affects the perceived haptic experience.
% }
%この研究の成果は、直接的な文脈を超えて、ツールを介した触覚研究とインターフェース設計におけるより広範な応用に対する貴重な洞察を提供します。私たちの結果は、実験設定と触覚インターフェースの両方で材料と構造の剛性特性を考慮することの重要性を強調しています。たとえば、ツールを介した触覚知覚に関する研究では、使用されるツールの剛性は、知覚される触覚体験に大きく影響するため、実験パラメータとして扱う必要があります。

% \add{
% In designing haptic feedback systems, these findings highlight the necessity of creating comprehensive haptic experiences by carefully factoring interface stiffness. By doing so, designers can ensure that the interface provides feedback that closely aligns with user expectations and enhances the realism of the virtual interactions.
% }
%触覚フィードバック システムを設計する場合、これらの成果は、インターフェースの剛性を慎重に考慮して包括的な触覚体験を作成する必要性を強調しています。そうすることで、設計者はインターフェースがユーザーの期待に密接に一致するフィードバックを提供し、仮想インタラクションのリアリティを高めることを保証できます。

\del{
As a future prospect, the current motor operated at a rated voltage of 12 V at 78 rpm and was selected for its ability to generate the highest torque among the available options. However, there is a trade-off between the torque and rotational speed (rpm). Although a higher rotational speed is required for real-time stiffness control, torque considerations are also critical.
}
%今後の展望として，現在使用しているモーターは定格電圧12V，78rpmであり，用意できるモーターの中で最もトルクが出せるモーターを選んだ．しかし，一般に，モータによって生成されるトルクとrpmの値（回転速度）はトレードオフであるため，リアルタイムで剛性を制御するには回転速度の速いモーターを使用する必要があるが，トルクについても考慮する必要がある．

%インターフェースの内部圧力
% \add{
% Demonstrating the relationship between the internal pressure and interface stiffness is essential. However, the structural limitations of the interface currently prevent the direct measurement of the internal pressure. Addressing these limitations and integrating such measurements into the model is a critical focus for future research.
% }
%内部圧力とインターフェースの剛性の関係を実証することは不可欠です。しかし、インターフェースの構造上の制限により、現在は内部圧力を直接測定することができません。これらの制限に対処し、そのような測定結果をモデルに統合することが、今後の研究の重要な焦点となります。
\del{
We believe that real-time control of the stiffness of the interface could enable the creation of illusions not only of an object's stiffness, but also of its shape. 
For example, when touching a rectangular object, we propose that gradually softening the interface as it contacts the edges can alter the perceived touch points, creating the illusion of interaction with a circular object.
}
%デバイスの剛性をリアルタイムで制御することができれば，物体の剛性だけでなく，形を錯覚させることもできると私達は考える．例えば，四角形の物体に対してデバイスで触り，四角形の端に触れるにつれてデバイスを軟らかくすることで触れることができる位置が変わり，円形の物体に触れているように錯覚させることもできるのではないかと考えている．

\section{Conclusion}

In this study, we investigated the haptic perception of objects via a stylus and, based on the findings, proposed a stylus-shaped interface equipped with joints that can control stiffness. 
We further investigated the effectiveness of this interface in providing haptic rendering of virtual objects.
%本研究では，スタイラスを介した物体の触覚知覚について調査し，その知見を基に動的に剛性を変更できるジョイントを備えたスタイラス型インターフェースを提案し，実際にデバイスがバーチャルオブジェクトの触覚フィードバックができるのか調査した．

First, we investigated whether it was possible to distinguish between the stiffness of the stylus and the object when the visual stimuli depicted either the stylus or the object being deformed, and the subject used the stylus to touch the object. The results showed that by using a soft stylus, subjects could either perceive the stylus as soft, depending on the visual stimulus, or experience the illusion that the object is soft. This confirms that the haptic perception of stiffness can be altered by manipulating visual stimuli.
%まず，棒と物体のどちらかが変形する視覚刺激を提示して棒を用いて物体に触れる際，棒と物体のどちらが軟らかいか判別できるか調査した．その結果，軟らかい棒を用いることで，視覚刺激の種類によって棒が軟らかいと認識させることも，物体が軟らかいと錯覚を起こすこともできる可能性を示した．このように，視覚刺激の操作によって，硬軟知覚の変容が可能であることを確認できた．

Next, based on haptic perception via a stylus, we propose a stylus-shaped interface that alters the sensation felt by the hand by controlling the stiffness of its joints.
The interface is equipped with a piston mechanism driven by a small motor and features a two-layer tube structure that mimics the McKibben-type artificial muscles. The stiffness of the interface was controlled by changing the rotational direction of the motor, allowing seamless transitions between softness and stiffness. This design results in a compact, grip-type haptic interface with a simple structure, overcoming the challenges encountered in previous research, such as the large interface size, complex mechanisms, limited application areas, and reliance on specialized materials.  We developed interfaces with joint lengths of 10, 15, and \SI{20}{mm} and measured their stiffness. The results showed that the interface could generate sufficient stiffness to create the illusion that a hard object felt soft.
%次にスタイラスを介した触覚知覚を基に，関節の剛性を変化させることで，手から感じる感覚を変化させるスタイラス型インターフェースを提案した．デバイスは，小型のモータを利用したピストン機構と，マッキベン型人工筋肉を模した2層のチューブ構造を備えることで，モータの回転方向の変化のみで硬軟を制御する．これにより，デバイスが大型になる・複雑な機構が必要・応用先が限定される・特殊な素材を用いるといった従来研究の課題に捕らわれない，簡素な構造且つ小型の把持型触覚デバイスを実現した．そして，関節長が10mm/15mm/20mmのデバイスを作成し，剛性を測定した．その結果，物体が軟らかいという錯覚を起こすための剛性をデバイスが十分に出力できることがわかった．

Finally, we conducted a user study to evaluate whether the interface could present stiffness that accurately corresponded to the material of the virtual object. 
The results indicated that a joint length of \SI{15}{mm} was sufficient to provide haptic rendering consistent with the perceived material properties of the virtual object.
%The results indicated that a joint length of \SI{15}{mm} was sufficient to provide haptic feedback aligned with the perceived material properties of the virtual object.
%最後にデバイスがバーチャルオブジェクトのマテリアルに適した剛性を提示できるのか調査するためのユーザスタディを行った．その結果，関節長が15ｍｍのデバイスでは，十分にバーチャルオブジェクトのマテリアルの触覚フィードバックが可能であることがわかった．
Our results encourage future
research to uncover the full potential of haptic feedback for
VR based on stiffness of stylus.

%\lipsum[1]%

% \section*{Supplemental Materials}
% \label{sec:supplemental_materials}

% Refer to the instructions for this section (\cref{sec:supplement_inst}).
% Below is an example you can follow that includes the actual supplemental material for this template:

% All supplemental materials are available on OSF at \url{https://doi.org/10.17605/OSF.IO/2NBSG}, released under a CC BY 4.0 license.
% In particular, they include (1) Excel files containing the data for and analyses for creating \cref{tab:vis_papers} and \cref{fig:vis_papers}, (2) figure images in multiple formats, and (3) a full version of this paper with all appendices.
% Our other code is intellectual property of a corporation---Starbucks Research---and there is no feasible way to share it publicly.

% \section*{Figure Credits}
% \label{sec:figure_credits}

% Refer to the instructions for this section (\cref{sec:figure_credits_inst}).
% Here are the actual figure credits for this template:

% \Cref{fig:teaser} image credit: Scott Miller / Special to the Vancouver Sun, January 22, 2009, page A6.

% \Cref{fig:vis_papers} is a partial recreation of Fig.\ 1 from \cite{Isenberg:2017:VMC}, which is in the public domain.

\acknowledgments{
This research was partially supported by JSPS Grant-in-Aid JP22K18424.
%The authors wish to thank A, B, and C. This work was supported in part by a grant from XYZ.
}

\bibliographystyle{abbrv-doi}

\bibliography{template}
\end{document}